# Latest news on zebra patterns

Gennady Chernov

**Abstract** The publications of the last three years concerning the study of the most intriguing fine structure in type IV solar radio bursts – zebra pattern (ZPs), are surveyed. The main attention is paid to new observations, irrespective of whether a paper includes detailed interpretation of an event or simply reports the beginning of operation of a new tool. The radiation mechanism of a ZP on a double plasma resonance (DPR) remains the most widespread and standard, though ten alternative mechanisms have been offered. However, in a number of works difficulties with the explanation of a complex zebra are noted, especially in combination with fiber bursts and spikes. Therefore, several papers in which description of the radiation mechanism of a ZP on the DPR is improved are considered in more detail. Without positional observations we have a great opportunity to follow the dynamics of flare processes using SDO/AIA images in several EUV lines. In the discussion, the debatable questions regarding the comparison of mechanisms associated with DPR in terms of the model of interaction of plasma waves with whistlers are illuminated.
**Keywords** Sun: flares · Sun: fine structure · Sun: microwave radiation · Sun: zebra pattern

## 1. INTRODUCTION

A zebra-pattern (ZP), in the form regular stripes in emission and absorption on the dynamic spectra of solar radio bursts, has already been studied during the fifth solar cycle. Basic observational properties are presented in a number of reviews and monographs (Slottje,1981; Kuijpers, 1975; Chernov, 2006, Chernov, 2011 ). Simultaneously with the observations, theoretical models were being developed, and currently more than ten mechanisms are offered. Most often in the literature the mechanism based on a double plasma resonance (DPR) is discussed (Kuijpers, 1975; Zheleznykov and Zlotnik, 1975a,b; Kuijpers (1980) ; Mollwo, 1983; 1988; Winglee and Dulk, 1986) which assumes that the upper hybrid frequency ($\omega_{UH}$) in the solar corona becomes a multiple of the electron-cyclotron frequency:

$$\omega_{UH} = (\omega^2_{Pe} + \omega^2_{Be})^{1/2} = s\omega_{Be} \qquad (1)$$

where $\omega_{Pe}$ is the electron plasma frequency, $\omega_{Be}$ is the electron cyclotron frequency, $s$ is the integer harmonic number, and usually $\omega_{Be} << \omega_{Pe}$.

This mechanism experiences a number of difficulties with the explanation of the dynamics of zebra stripes and some thin effects (sharp changes of frequency drift, a large number of stripes, frequency splitting of stripes, superfine millisecond structure); therefore works on its improvement began to appear (Karlický et al. 2001; LaBelle et al. 2003; Kuznetsov and Tsap, 2007). The theory of LaBelle et al. 2003 is based on the emission of auroral choruses (magnetospheric bursts) via the escape of the Z mode captured by regular plasma density inhomogeneities. Kuznetsov and Tsap (2007) assumed that the velocity distribution function of hot electrons within the loss cone can be described by a power law with an exponent of 8–10. In this case, a fairly deep modulation can be achieved.

An alternative mechanism for ZP was proposed by Chernov (1976; 1990): the coalescence of plasma waves (*l*) with whistlers (*w*), $l + w \rightarrow t$ (Kuijpers, 1975), but an unified model in which the formation of ZP was attributed to the oblique propagation of whistlers, while the formation of stripes with a stable negative frequency drift (the fiber bursts) was explained by the ducted propagation of waves along a magnetic trap. This model explains occasionally observed transformation of the ZP stripes into fibers and vice versa.

---

---

G. Chernov
Key Laboratory of Solar Activity, National Astronomical Observatories, Chinese Academy of Sciences, Beijing 100012, China
Pushkov Institute of Terrestrial Magnetism, Ionosphere and Radio Wave Propagation, Russian Academy of Sciences (IZMIRAN), Troitsk, Moscow 142190, Russia   e-mail: gchernov@izmiran.ru

However, the certain boom of new models proceeds. In several works the formation of ZP stripes is explained by radio wave diffraction on the heterogeneities in the corona (Laptukhov and Chernov, 2006; Barta and Karlicky, 2006), or by the interference of straight and reflected from the heterogeneities rays (Ledenev, Yan, and Fu, 2006). In these cases the number of harmonics does not depend on the ratio of the plasma frequency to the gyrofrequency in the source. The formation of ZP stripes due to radio wave propagation through the coronal heterogeneities can be recognized as the most natural mechanism of ZP.

It is important to note that the model with the whistlers successfully explains the zigzags of stripes and their splitting, and also the variations in the frequency drift of stripes synchronously with the spatial drift of the sources of radio emission (Chernov, 2006). Since each new phenomenon provides its uncommon parameters of fine structure, the entire variety of parameters does not succeed in the statistical systematizing (e.g. see Tan et al. 2014). Below primary attention is given to the analysis of separate phenomena. Just such a situation stimulates many authors to elaborate on new mechanisms.

Only positional observations could help to find out where the ZP stripes form (during the excitation of waves in the source or in the course of their further propagation).

During the last several years some new varieties of ZP have been recorded. In the present paper an attempt is made to estimate when the positional observations can be the determining factor for the selection of the radio emission mechanism.

We also try to evaluate what model most adequately describes the new observational data and to find out where the ZP stripes form (during the excitation of waves in the source or in the course of their further propagation). Calculations show that the DPR-based mechanism fails to describe the generation of a large number of ZP stripes in any coronal plasma model. Karlicky and Yasnov, 2015 also examined some other unsolved problems or difficulties in the DPR model in detail.

Here, it is shown that the new varieties of ZP succeed in explaining these phenomena within the framework of known mechanisms by taking into account the special features of plasma parameters and fast particles in the source. On the other hand, the formation of ZP stripes due to radio wave propagation through the coronal heterogeneities can be recognized as the most natural mechanism of ZP. The mechanism related to the excitation of discrete eigenmodes of the periodically nonuniform plasma (Laptukhov and Chernov, 2006; 2009; 2012) can yield the observed number of harmonics. However, in this case, only the possibility of generating harmonics in a one-dimensional stationary problem is considered, i.e., the frequency dynamics of stripes is not analyzed.

During the last several years many new events with ZP have been recorded. Now, it is necessary to estimate the possibility of their interpretation by taking into account all known models of ZPs, and our main goal is to cover of results of all recent publications about a ZP.

**2. SOME FRAGMENTS OF NEW ORIGINAL OBSERVATION**

The measurement of positions and sizes of radio sources in the observations of the fine structure of solar radio bursts is the determining factor for the selection of the radio emission mechanism. The identical parameters of the radio sources for zebra-structure and fiber bursts will confirm the united mechanism for both structures.

The new examined events show that the zebra-structure and fiber bursts can appear almost simultaneously or consecutively in the microwave, decimeter and meter wave bands.

2.1. FASR/FST observations of ZP on December 14, 2006

The most cited and used model for ZP is mechanism at DPR, especially after the paper of Chen Bastian et al. (2011) where the authors allegedly confirm this mechanism using positional observation by the Frequency-Agile Solar Radiotelescope Subsystem Testbed (FASR, FST).

*Observation.* **They present the first interferometric observation of a radio burst with ZP in the decimeter range in the event on December 14, 2006 with simultaneous high spectral (≈1 MHz) and high time (20 ms) resolution (Figure 1). By using Owens Valley Solar Array (OVSA) to calibrate the FST, the source position of the zebra pattern can be located on the solar disk. The location of the ZP** emission centroid was defined by the intersection of the three interferometric fringes of FST. The interferometric phase embodies the spatial information of the radiation source. They found a spatial drift of ZP source of 15.6 ± 6″.5 from NE to SW on the solar disk, corresponding to a projected drift velocity of $2.5 \pm 1.0 \cdot 10^9$ cm s$^{-1}$ (≈0.1$c$). Unfortunately, Chen, Bastian et al. (2011) considered averages of the phases "along" the on- and off-stripe positions (in time), to reveal the spatial distribution of the ZP source over frequency.

*Comments to the interpretation.* The authors conclude that the zebra burst is consistent with a DPR model, and they reject the alternative whistler model using an estimation of the tangent angle between whistler trajectory and solar surface (where a moving radio source was observed) by two method.

In this connection, it is necessary to note that the model with whistlers was there mistakenly rejected. In the whistler model, when we estimate $V_{gr} = 2.5 \times 10^9$ cm s$^{-1}$ , that's value in the quasi-longitudinal propagation (along the magnetic trap, it is not $V_{proj}$ as authors used ). And $V_{proj} = V_{gr} \cos \alpha_1$ , then for big whistler angle propagation (cos $\alpha_1$~0.1) you could receive the same value of tan $\alpha_1$ , as for tan $\alpha_2$ (see Figure 2). So, no problem with the whistler model, in which the spatial drift of radio source can be actually connected with the projection of the group velocity of whistlers. In such a case the assumed relation of the spatial drift of the radio source and frequency drift of stripes on the spectrum with a change in the speed of fast particles drops off.

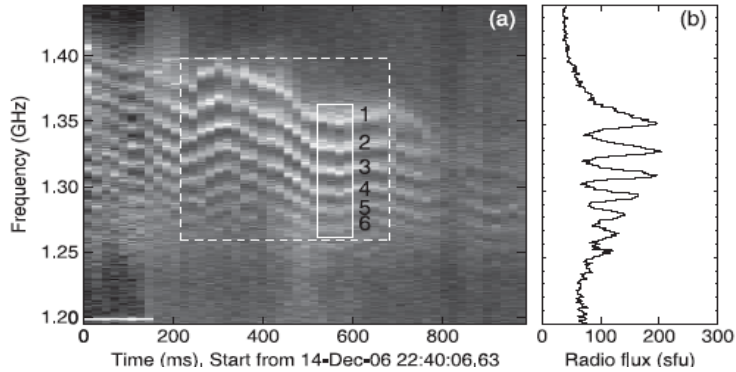

**Figure 1** (a) A zebra-pattern structure observed at around 22:40 UT on 2006 December 14. Six successive strong stripes with decreasing frequencies are marked by numbers 1–6. (b) Frequency profile of zebra-pattern structure, averaged in the time denoted by the small solid box (fragment of Fig. 6 from Chen, Bastian et al. 2011).

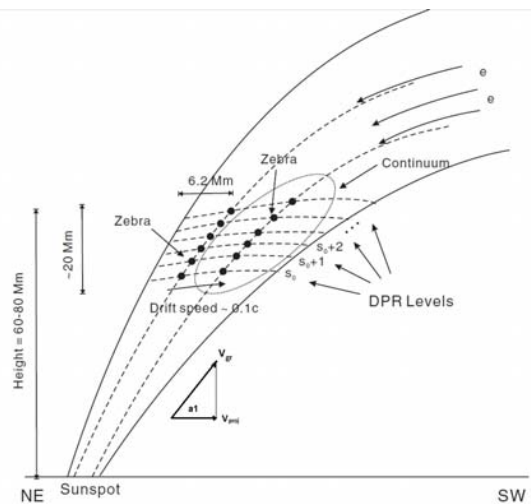

**Figure 2** Simplified source model (Fig. 14 from Chen, Bastian et al. (2011)), supplemented by the vector of the group velocity.

The frequency drift of the of zebra- stripes three times changed the sign during one second (Figure 1), which highly improbably can be connected both with the change in speeds of fast particles and with rapid changes in the magnetic field. The high speed of spatial drift of the zebra sources (~0.1 c) can be related with more high density gradient across the magnetic trap.

Then, Chen, Bastian et al. (2011) used convenient idealistic models for plasma density and magnetic field with exponential dependence, which allows to receive the necessary ratio of scales heights for density and magnetic field $L_n/L_B$ = 4.4. In the decimeter range an attempt at the determination of the displacement of source during one stripe of zebra in phenomenon on December 14, 2006 with the aid of the system FASR did not succeed because of the insufficient time resolution, ~ 20 ms approximately in the same time interval.

The height of source was determined by the indirect method: by the extrapolation of the magnetic field above the active region in combination with the arbitrarily selected density model. Thus, it is impossible to recognize the selection of model for ZP in this work unambiguous even for one considered event. Additional similar observations for more detailed consideration of behavior of a radio source of one stripe of the ZP are required.

In the whistler model, radio sources of fiber bursts and ZP should be really moving, and the spatial drift of ZP stripes should change synchronously with changes of frequency drift in the dynamical spectrum. This occurs in accordance with a change in the group velocity of whistlers as a result of the quasi-linear diffusion of the latter on fast particles (Chernov, 1990). In the DPR model the ZP source must be rather stationary.

2.2. Several papers were devoted to analysis of some selected events
2.2.1. The September 17, 2002 event
It is known that ZP is rarely observed at frequencies higher than 5 GHz.
From new publications, Altyntsev et al. (2011) discuss the event on September 17, 2002 as a zebra-like one (Figure 3). Although it is difficult to rank this image as typical zebra, Altyntsev et al. discuss it in frame of DPR model, together with several other events, among which one was May 29, 2003 event.

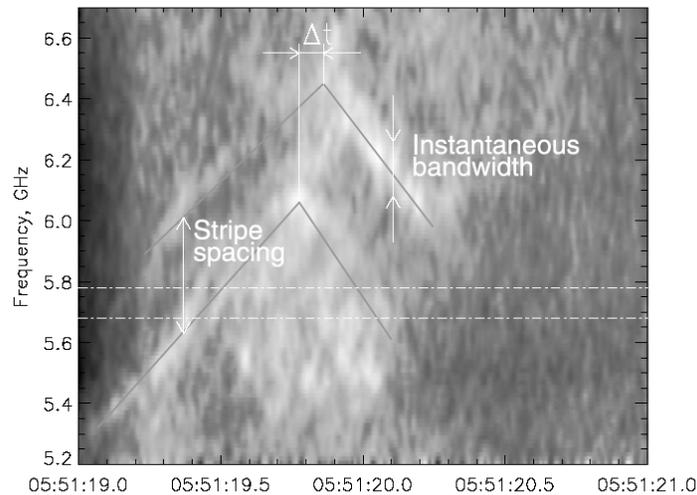

**Figure 3** The dynamic spectrum of the zebra structure observed on 17 September 2002. White means increased emission. The zebra stripes are indicated by the gray lines. The horizontal lines mark the boundaries of the SSRT receiver bands (from Altyntsev et al. (2011).

2.2.2. The May 29, 2003 event

This event was analyzed also by Chernov et al. (2012a) in more details (Figure 4). Dynamic spectra in Figure 4 demonstrate that all the emission comprises fine-structure details in the form of pulses (spikes). Their duration is limited to one pixel; i.e., it is at the limit of the 5 ms temporal resolution (rarely to two pixels, ~10 ms). They occupy the frequency band in the spectrum of ~70 to 100 MHz. Their band and duration might be below the resolution limit. The continual background burst also exhibits the spike structure. Groups of spikes create various fine-structure forms: fast and slow-drift fibers. The brightest (most intense) ones form different zebra stripes.

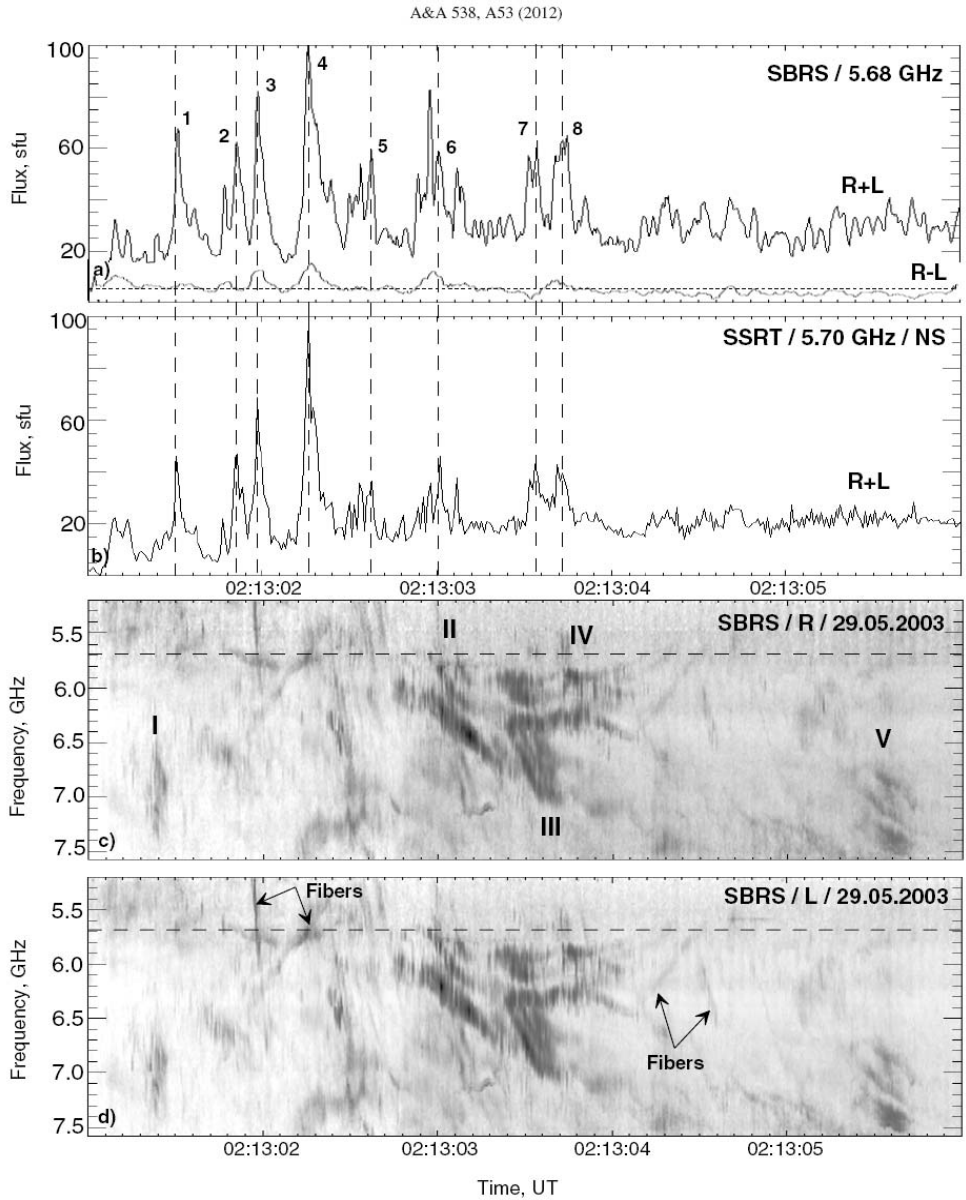

**Figure 4** Comparison of the time profiles recorded at Solar Broadband Radio Spectrometer (SBRS) **a)** and SSRT **b)** and the dynamic spectrum **(c)**, **(d)** registered by the SBRS spectro-polarimeter. The dark details correspond to the increased emission. The horizontal dashed line **c)** denotes the SSRT receiving frequency at 5.7 GHz. The calibration was performed using the SSRT-registered background burst. Arabic numerals (**a)**, **b)**, dashed vertical lines) mark the maximum moments of subsecond pulses recorded by the SSRT (from Chernov et al. (2012a).

Positions of radio bursts were obtained by the Siberian Solar Radio Telescope (SSRT) (5.7 GHz) and Nobeyama radioheliograph (NoRH) (17 GHz) (Figure 5).

This event may be used as an example to demonstrate the possible difficulties that the mechanism at DPR faces for explaining the zebra-structure with sporadic stripes. Originally, it was developed on the condition that there are always DPR levels in the source, and only the presence of fast particles with loss-cone or ring (DGH) velocity distribution provides the excitation of the regular zebra structure of different modulation depths (depending on the particle energy spectrum, Kuznetsov and Tsap (2007)). The presence of such particles may be considered obvious, so that the absence of zebra-structure at the beginning of the burst suggests there are no DPR levels.

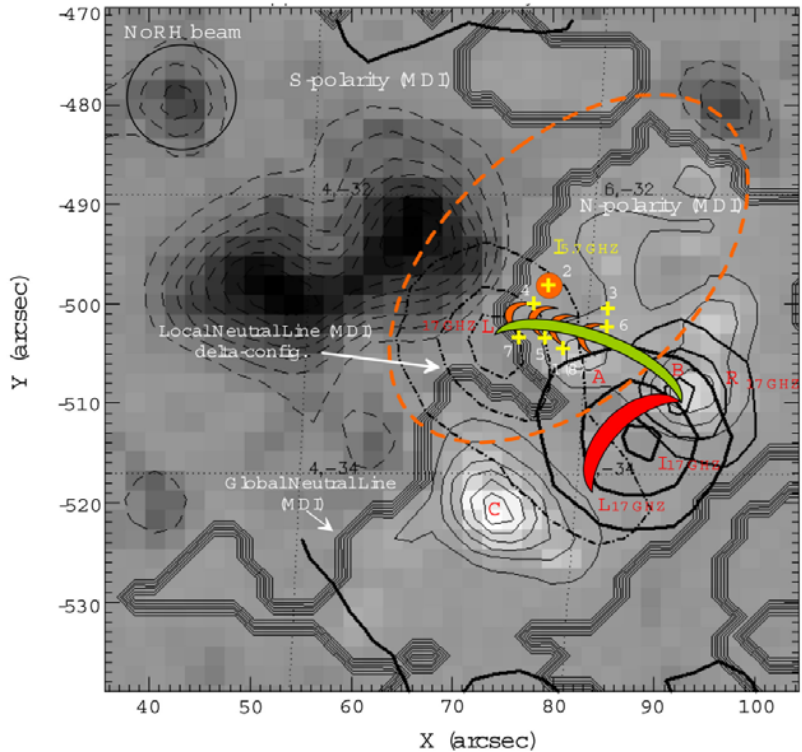

**Figure 5** The positions of radio sources at 5.7 GHz (SSRT) and 17 GHz (NoRH) superimposed on the MDI magnetogram. The orange dashed line indicates the position of the left polarized local source at 5.7 GHz at 02:17:54.8 UT. The dash-and-dot line shows the position of the left polarized background source of the burst at 17 GHz (extended southwestward) at 02:13:00 UT. The straight yellow crosses are the positions of spike sources 1–8 corresponding ones in Figure 4 (from Chernov et al. (2012a).

DPR levels cannot quickly appear and disappear in the corona. Processes of emergence of new magnetic fluxes and formation of new loops are very slow. Changes in height scales of density and magnetic fields may be associated only with flare ejections and shocks. With the ejection velocity taken as Alfvén velocity ($\approx 1000$ km s$^{-1}$), the DPR levels (in view of the magnetic field line freezing-in) may shift the flare loop by half (green in Figure 5) with sizes of ~10″ for approximately 7 s. This time exceeds the lifetime of zebra-structure stripes by an order of magnitude. The sporadic nature of ZS stripes, the wave-like drift of two stripes, their splitting, and the presence of isolated fibers with forward and reverse drifts rule out the existence of any regular DPR levels. Moreover, the validity of the DPR conditions in fine flare loops along which density and magnetic field strength vary only slightly is in doubt (Aschwanden 2004).

At the same time, all the effects can be naturally explained in the joint model of zebra-structure and fibers when plasma waves interact with whistlers (Chernov 2006).

Then, estimations carried out in Chernov et al. (2012a) show that propagation time of whistlers along flare loops corresponds to life time of zebra stripes.

The sporadic nature of the fine structure suggests multiple pulsating acceleration of fast particles. However, the superfine spike structure is not necessarily determined by a pulsating acceleration as in Kuznetsov (2007). It is much more probable that the spike nature of the entire emission is formed by the pulsating mechanism of emission. ZP is generated by periodic whistler packets filling a magnetic trap. But whistlers should occupy the entire radio source and propagate in different directions. Spikes might therefore be related to the pulsating interaction of whistlers with ion-sound waves and subsequent coalescence with plasma waves. This mechanism is discussed in Chernov et al. (2001). In the microwave range it may be more effective because near the reconnection region the presence of nonisothermic plasma ($T_e >> T_i$) is much more likely as a condition for ion-sound excitation.

2.2.3. The event on February15, 2011

Tan et al. (2012) analyzed ZP in the first X- class flare in the solar cycle 24 on February15, 2011. Fig. 1 of this work depicts the temporary profiles of the fluxes of radio emission in a number of frequencies and the profiles of soft X-ray GOES 4 and 8. X-ray flare began approximately on 15 min earlier than radio bursts. The temperature of plasma in the X-ray sources was very high, on the order of 20 MK. Three ZS stripes in the microwave range appeared first at frequencies 6.4 – 7.0 GHz in the left-handed polarization during the rising phase of the first flare brightening (ZP1, Figure 6). Then, at 11 min ZP2 appeared on the decrease of the second brightening in the range 2.6 – 2.75 GHz (shown here in Figure 7) also in the left-handed polarization. Further, still at 9.5 min on the decrease of the third brightening in the decimeter range 1035 – 1050 MHz, ZP3 was registered by SBRS/Yunnan (see Figure 8) but already with the moderate right-handed polarization. Thus, in the prolonged event ( 45 min) the ZS appeared only three times for 3-second intervals. By scrutinizing the current prevalent theoretical models of ZP structure generations and comparing their estimated magnetic field strengths in the corresponding source regions, Tan et al. (2012) suggest that the double plasma resonance model is the most probable one for explaining the formation of microwave ZPs.

2.2.3.1. Chernov et al. 2015 focus attention on a number of the aspects, not examined in the work of Tan et al. (2012). In particular, zebra stripes were observed also in the meter range, according to the Culgoora spectrograph in the range 200 - 300 MHz (see Fig. 10 in Chernov et al. 2015). The type II burst is also visible, which was begun near 01:50 UT. The beginning of the corresponding CME comes approximately on 01:30 UT, that coincides with the beginning of the X-ray burst. Therefore the type II burst was most likely caused by piston shock wave.

In Figure 7 we see fast pulsations in 2.7-2.9 GHz band, from high frequencies of ZP. After approximately 2.5 min. specifically, in this range fiber bursts appeared with the intermediate frequency drift *df/dt* of ≈ 350 MHz s$^{-1}$, approximately with the same period (see Figure 11 in Chernov et al. 2015). Subsequently, during five minutes the initial frequencies of fibers increased approximately on 100 MHz together with a new series of pulsations with the irregular period.

If the onsets of three intervals ZP1, ZP2 and ZP3 to connect with the appropriate flare brightenings and the ejections in the SDO/AIA of 171 Å images, then they lie down in the different sections of AR. In Figure 12 in Chernov et al. 2015 it is shown that ZP1 and ZP2 lie down on the loops above the preceding spot of northern magnetic polarity, and ZP3 above the tailed spot of southern polarity. Therefore in all three cases emission can be connected with the ordinary wave.

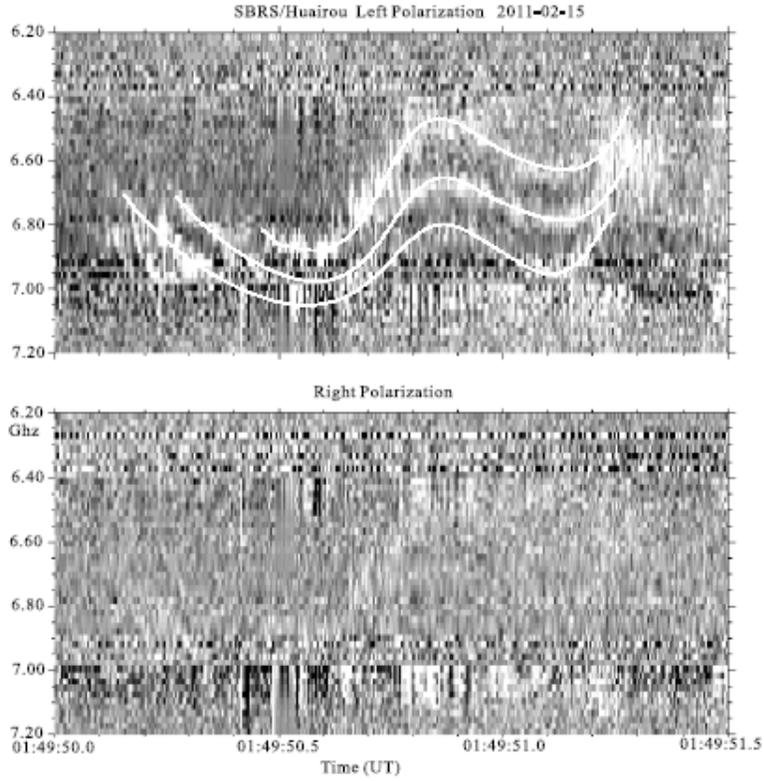

**Figure 6** Three ZS stripes in the range 6.4 – 7.0 GHz in the Left polarization channel of SBRS/Huairou at rising phase of the first flare brightenning (ZP1 in Tan et al. 2012). The white dashed curves outshine the zebra stripes.

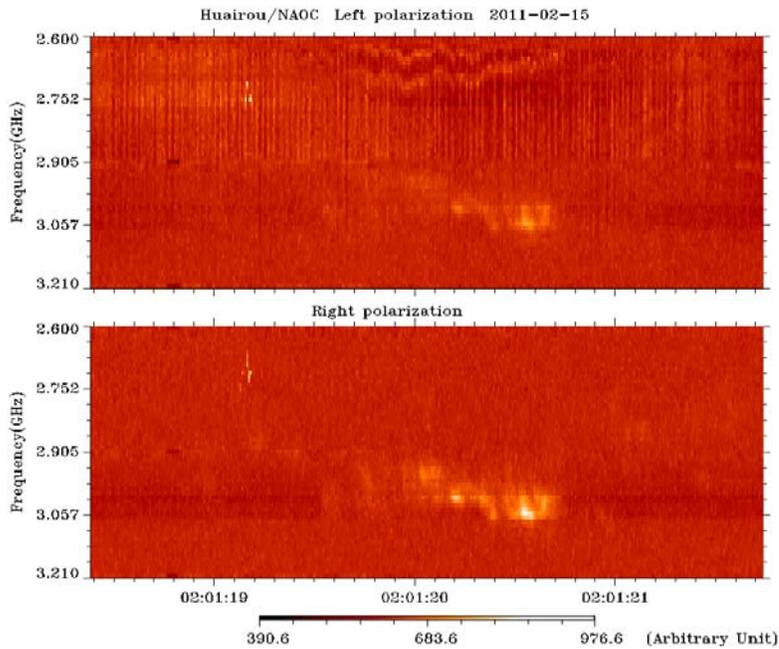

**Figure 7** Two stripes of ZS superimposed on fast pulsations with the period of ~ 30 ms.

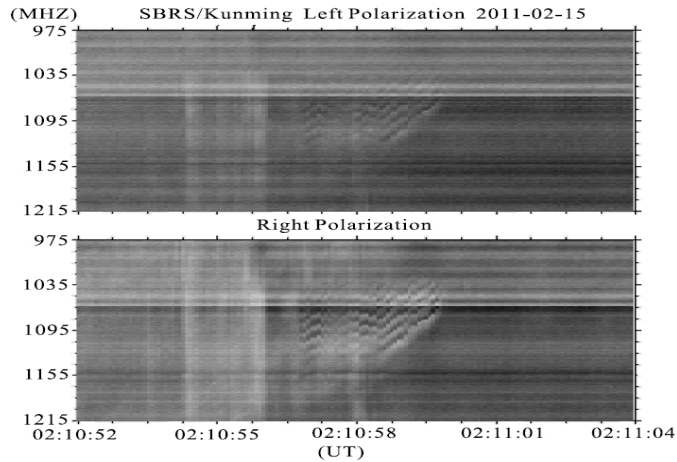

**Figure 8** Zebra- structure in the decimeter range 1.04 – 1.13 GHz in right polarization registered by the SBRS/Yunnan (Tan et al. 2012).

2.2.4. Fast pulsation in the event on August 9, 2011

Tan B. and Tan Ch. (2012) analyze peculiar microwave quasi-periodic pulsations (QPP) accompanying a hard X-ray (HXR) QPP of about 20 s duration observed just before the maximum of an X6.9 solar flare on August 9, 2011 at 2.60–3.80 GHz by the Chinese Solar Broadband Radio Spectrometer (SBRS/Huairou). The most interesting aspect is that the microwave QPP consists of millisecond timescale superfine structures and zebra stripes at low frequency part of QPP (Figure 9). Each microwave QPP pulse is made up of clusters of millisecond spike bursts or narrow-band type III bursts. The physical analysis indicates that the energetic electrons accelerating from a large-scale highly dynamic magnetic reconnecting current sheet above the flaring loop propagate downward, impact the flaring plasma loop, and produce HXR bursts. The tearing-mode (TM) oscillations in the current sheet modulate HXR emission and generate HXR QPP; the energetic electrons propagating downward produce Langmuir turbulence and plasma waves, resulting in plasma emission. The modulation of TM oscillation on the plasma emission in the current-carrying plasma loop may generate microwave QPP. The TM instability produces magnetic islands in the loop. Each X-point will be a small reconnection site and will accelerate the ambient electrons. These accelerated electrons impact the ambient plasma and trigger the millisecond spike clusters or the group of type III bursts. Possibly, each millisecond spike burst or type III burst is one of the elementary bursts (EBs). A large number of such EB clusters form an intense flaring microwave burst. This original interpretation is illustrated by schematic in the Figure 10.

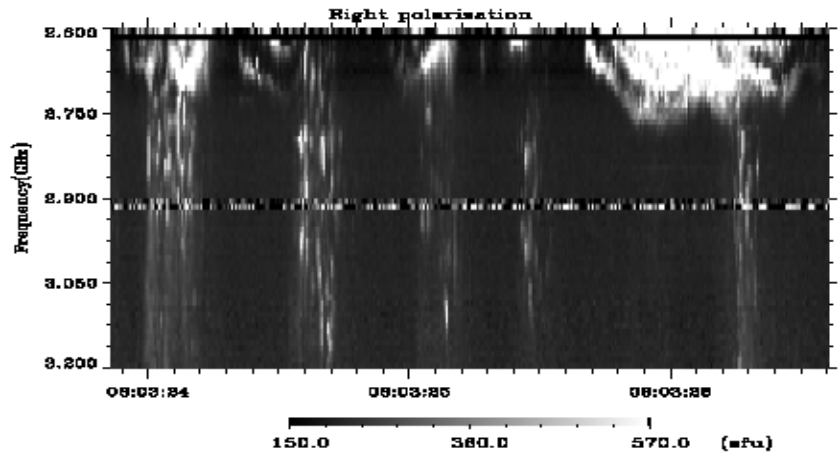

**Figure 9** Three stripes of ZS in the LF edge of pulsations with millisecond superfine structure (from Tan B. and Tan Ch. (2012).

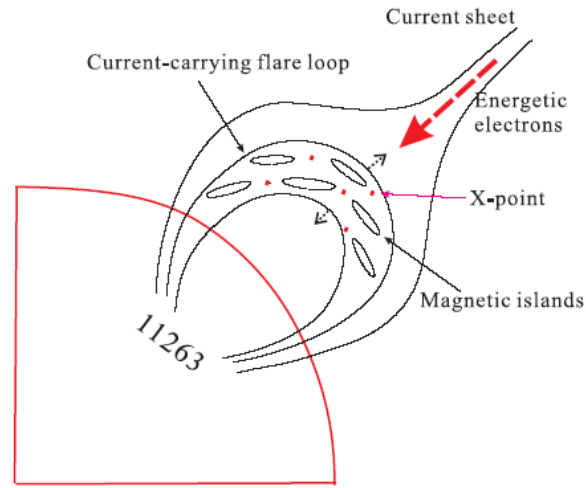

**Figure 10** Schematic showing the processes of electron acceleration above and in the flare loop, propagation, and the formation and distribution of magnetic islands in the current-carrying flare loop. It is possible only to add that a source of ZP is probably located at the top of current-carrying flare loop (from Tan B. and Tan Ch. (2012) .

2.2.5. The decay phase of the event on December 1, 2004
Huang and Tan (2012) made a detailed analysis of the microwave FSs during the decay phase of the M1.1 flare on December 1, 2004 by using the microwave observations of SBRS/Huairou ranging from 1.10 to 7.60 GHz and the HXR observations from RHESSI.

In the microwave spectra, they have identified stripe-like bursts such as lace bursts, fiber structures, zebra patterns (ZPs), and quasi-periodic pulsations. They also have detected short narrowband bursts such as dots, type III, and spikes. The lace bursts had rarely been reported, but in this event they are observed to occur frequently in the decay phase of the flare. The similarity between 25 and 50 keV HXR light curve and microwave time profiles at 1.10–1.34 GHz suggests that these microwave FSs are related to the properties of electron acceleration. The electron velocity inferred from the frequency drift rates in short narrowband bursts is in the range of 0.13 c –0.53 c and the corresponding energy is about 10–85 keV, which is close to the energy of HXR-emitting electrons. From the Alfven soliton model of fiber structures, the double plasma resonance model of ZPs, and the Bernstein model of the lace bursts, the authors derived a similar magnetic field strength in the range of 60–70 G. Additionally, the physical conditions of the source regions such as height, width, and velocity are estimated.

The lace bursts are the most obvious fine structure in the decay phase of the M1.1 flare on December 1, 2004 (Figure 11). According to Karlicky et al. (2001), the lace bursts may be generated by DPR mechanism in turbulent plasmas associated with the plasma outflows from the magnetic reconnection sites. In this case, naturally, the authors in no way can explain the sharp jumps of frequency drift, since actually the precise mechanism is unknown (turbulence). Nevertheless, the authors determine the magnetic field strength, using the Bernstein model. And it is no matter how strange, they obtained the same strength as well as obtained with fibers bursts and ZP. At least, such an agreement does not prove the correctness of their selection of the Bernstein model.

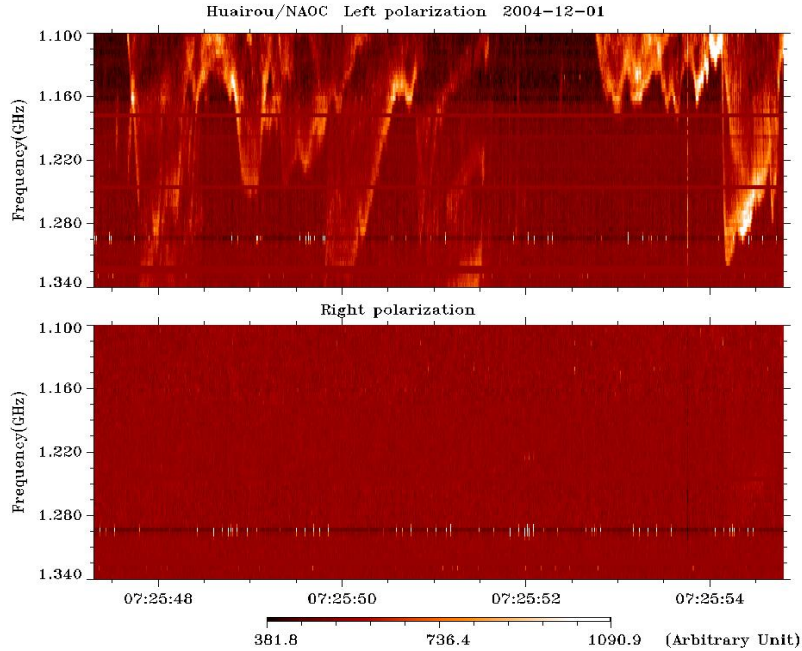

**Figure 11** Dynamic spectrum of lace bursts in decay phase of the M1.1 flare event observed at SBRS/Huairou at 07:25:48–07:25:55 UT on 2004 December 1 (Fig. 3 from Huang and Tan, 2012).

2.2.6. Quasi-periodic wiggles of ZP stripes
In Yu, Nakariakov et al. (2013) quasi-periodic wiggles of microwave zebra pattern (ZP) structures with periods ranging from about 0.5 s to 1.5 s are found in an X-class solar flare on December 13, 2006 at the 2.6–3.8 GHz with the Chinese Solar Broadband Radio Spectrometer (SBRS/Huairou). Periodogram and correlation analysis show that the wiggles have two–three significant periodicities and are almost in phase between stripes at different frequencies (Figure 12). The Alfven speed estimated from the ZP structures is about 700 km s$^{-1}$. They have found the spatial size of the wave-guiding plasma structure to be about 1 Mm with a detected period of about 1 s. This suggests that the ZP wiggles can be associated with the fast magneto-acoustic oscillations in the active flaring region. The lack of a significant phase shift between wiggles of different stripes suggests that the ZP wiggles are caused by a standing sausage oscillation.

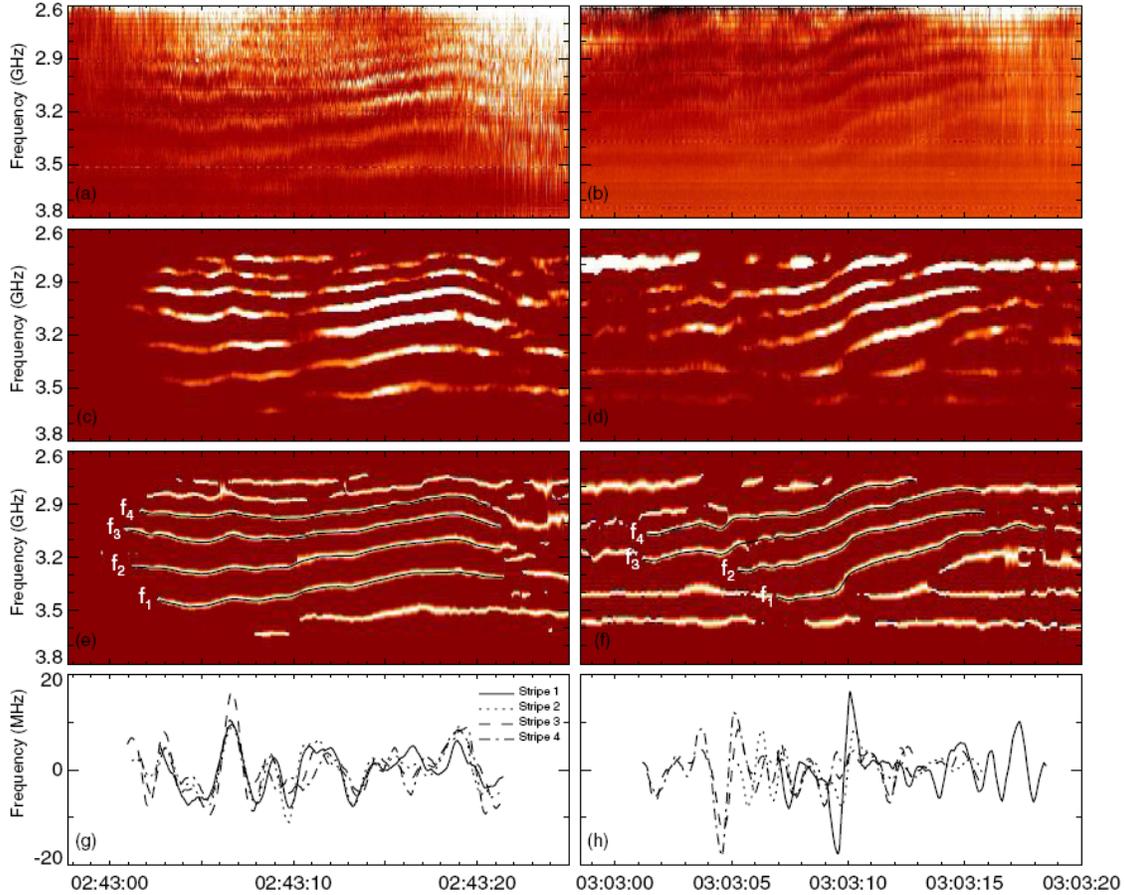

**Figure 12** Zebra pattern structures on 2006 December 13 observed by SBRS/Huairou at 2.6–3.8 GHz and the illustration of the processes of extracting the zebra pattern stripes. Panels (a, b): raw spectrograms at 2.6–3.8 GHz on LHCP; (c, d) high contrast images; (e, f) rescaled images with the extracted stripes superposed; and (g, h) detrended selected stripes frequency $f_N$ (from Yu, Nakariakov et al. (2013).

2.2.7. STATISTICS
Tan et al. (2014a) tried to carry out statistical comparison of models among which the following is chosen:
(1) Bernstein wave (BW) model;
(2) Whistler wave (WW) model;
(3) Double plasma resonance (DPR) model;
(4) Propagating model.

The microwave zebra pattern (ZP) is the most interesting, intriguing, and complex spectral structure frequently observed in solar flares. A comprehensive statistical study will certainly help us to understand the formation mechanism, which is not exactly clear now. Tan et al. (2014a) present a comprehensive statistical analysis of a big sample with 202 ZP events collected from observations at the Chinese Solar Broadband Radio Spectrometer at Huairou and the Ondřejov Radiospectrograph in the Czech Republic at frequencies of 1.00–7.60 GHz from 2000 to 2013. After investigating the parameter properties of ZPs, such as the occurrence in flare phase, frequency range, polarization degree, duration, etc., they have found that the variation of zebra stripe frequency separation with respect to frequency is the best indicator for a physical classification of ZPs. Microwave ZPs can be classified into three types: equidistant ZPs, variable-distant ZPs, and growing-distant ZPs, (Figure 13) possibly corresponding to mechanisms of the Bernstein wave model, whistler wave model, and double plasma resonance model, respectively. This statistical classification may help us to clarify the controversies between the existing various theoretical models and understand the physical processes in the source regions.

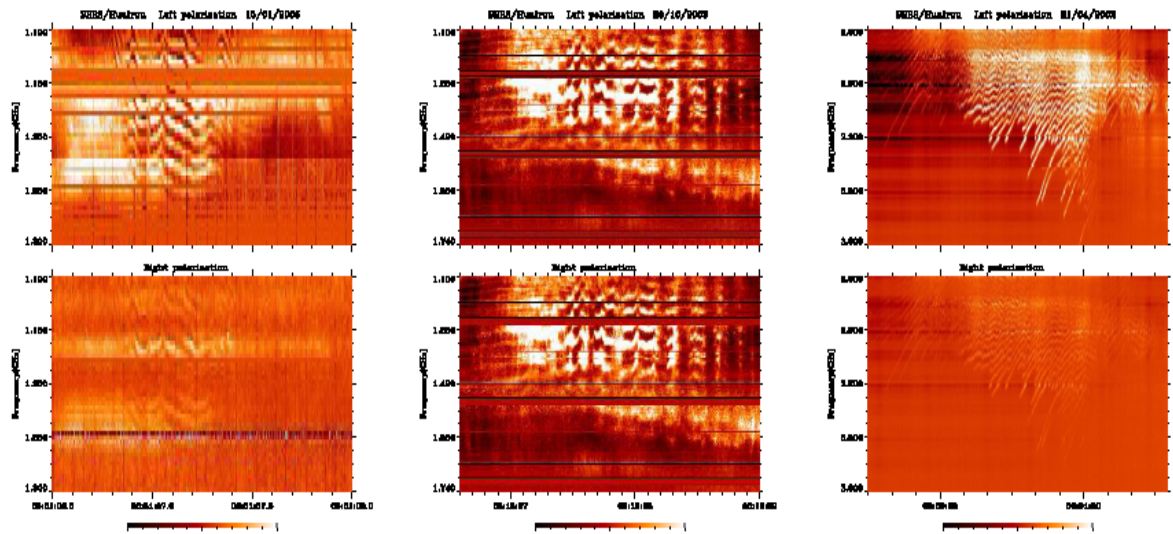

**Figure 13** Spectrograms with different ZP separation; (left)- typical equidistant with constant separation (BM), (middle)- variable-distant ZP (WW) with varying separation; and (right)- growing-distant with rising separation (DPR).

However, a problem remains, connected, first of all, with the ambiguous determination of frequency separation from the observations. Secondly, in the whistler model the frequency separation also increases with frequency in accordance with increasing of cyclotron frequency with plasma frequency.

2.2.8. The ratio $L_N/L_B$.
Yu, Yan, and Tan (2012) investigated the variations of 74 microwave ZP structures observed by the Chinese Solar Broadband Radio Spectrometer (SBRS/Huairou) at 2.6–3.8 GHz in nine solar flares. They used the DPR model with exponentially decrease of both the magnetic field strength ($B$) and the plasma density ($n_e$) with height ($h$) along the flux tube in the corona.

They found that the ratio between the plasma density scale height $L_N$ and the magnetic field scale height $L_B$ in emission sources displays a tendency to decrease during the flaring processes. The ratio $L_N/L_B$ is about 3–5 before the maximum of flares. It decreases to about 2 after the maximum. The variation of $L_N/L_B$ during the flaring process is most likely due to topological changes of the magnetic field in the flaring source region, and the stepwise decrease of $L_N/L_B$ possibly reflects the magnetic field relaxation relative to the plasma density when the flaring energy is released.

The ratio $L_N/L_B$ plays a significant role in the DPR model (Zheleznyakov & Zlotnik 1975). However, the absolute values of $L_N$ and $L_B$ are even more important. The authors discuss a contradiction with DPR model in April 21, 2002 event. Their estimations of ratio $L_N/L_B$ =2, magnetic field $B$ =19.5 G , and plasma $\beta \approx 1.4$ are a major obstacle for the DPR model to produce 34 ZP stripes simultaneously in the range 2.6–3.8 GHz in this event. And they proposed whistler model to be responsible for the formation of the multi-striped ZP structures in this event, as the ratio $L_N/L_B$ is not a significant parameter in this model, and the absolute value of magnetic field $B$ of $\approx$ 71.5 G is more important.

2.2.9. ZP during magnetic reconnection (Chernov et al. 2015)
Without positional observations we have a great opportunity to follow the dynamics of flare processes using SDO/AIA images in several EUV lines (Chernov et al. 2015). The event on February 24, 2011 is remarkable, as the zebra-structure at frequencies of 2.6 – 3.8 GHz was not polarized and it appeared during the magnetic reconnection observed by SDO/AIA 171 Å in this limb flare. (see Figures 2a and 6 in Chernov et al. 2015).

2.2.10. Rope-like fibers in the event on February 12, 2010

Solar radio emission records received at the IZMIRAN spectrograph (25–270 MHz) during the solar flare event of February 12, 2010 are analyzed by Chernov et. al. (2014). Different fine structures were observed in three large groups of type III bursts against a low continuum. According to data from the Nançay radioheliograph, sources of all three groups of bursts were located in one active region, 11046, and their emissions were accompanied by soft X- ray bursts (GOES satellite): C7.9 at 07:21 UT, B9.6 at 09:40 UT, and M8.3 at 11:25 UT. After the first group of bursts, classical fiber bursts were observed in combination with reverse-drift fiber bursts with unusual arc drift. After the third (the most powerful) group, stable second-length pulsations and slow drift fiber bursts were observed, the instantaneous frequency bands of which were an order of magnitude larger than the frequency band of classical fiber bursts, and the frequency drift was several times lower. More complex fiber bursts were observed in the weakest group in the time range 09:40:39–09:42:00 UT (Figure 14). They were narrow-band (~0.5 MHz) fiber bursts, periodically recurring in a narrow frequency band (5–6 MHz) during several seconds (rope – like fibers). The presence of many chaotically drifting ensembles of fibers, crossing and superimposing on one another, is a feature of this event. It is assumed that occurrence of these structures can be connected with the existence of many small shock fronts behind the leading edge of a coronal mass ejection.

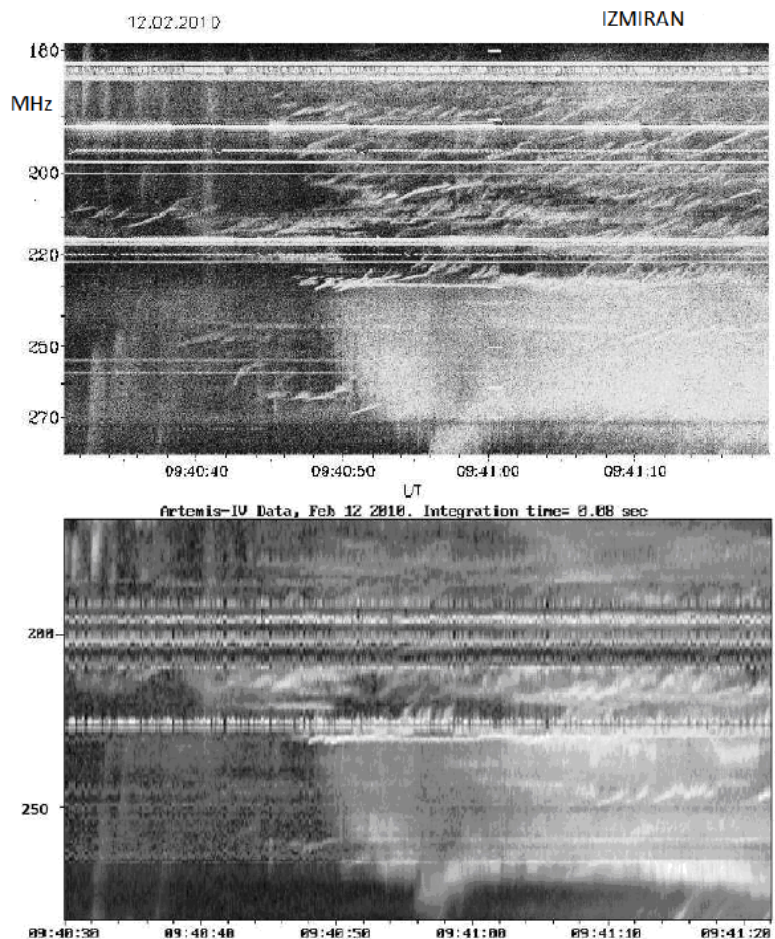

**Figure 14** Segments of dynamics spectra with fiber ropes in the 180–280 MHz range measured simultaneously on IZMIRAN and ARTEMIS_IV (Greece) spectrographs at February 12, 2010 (from Chernov et al. 2014b).

2.2.11. Flare evolution and polarization changes of ZP in the event April 11, 2013

2.2.11. Polarization changes of ZP in the event on April 11, 2011
In a M6.5 flare on April 11, 2013, solar radio spectral fine structures were observed for the first time simultaneously by several radio instruments at four different observatories. (Chernov et al. 2016). The fine structures include microwave zebra patterns (ZP) fast pulsations, and fibers. They were observed during the flare brightening located at the tops of a loop arcade as shown in SDO/AIA images.

At the beginning of the flare impulsive phase, a strong narrowband ZP burst occurred with a moderate left-handed circular polarization (Figure 15, see also Tan et al. 2014b)). Then a series of pulsations and ZPs were observed almost in an unpolarized emission. After 07:00 UT a ZP appeared with a moderate right-handed polarization (Figure 16). In the flare decay phase (at about of 07:25 UT), ZP and fiber bursts become strongly right-hand polarized. Combining magnetograms observed by the SDO Helioseismic and magnetic Imager (HMI) with the homologous assumption of EUV flare brightenings and ZP bursts, we deduced that the observed ZPs correspond to the ordinary radio emission mode.

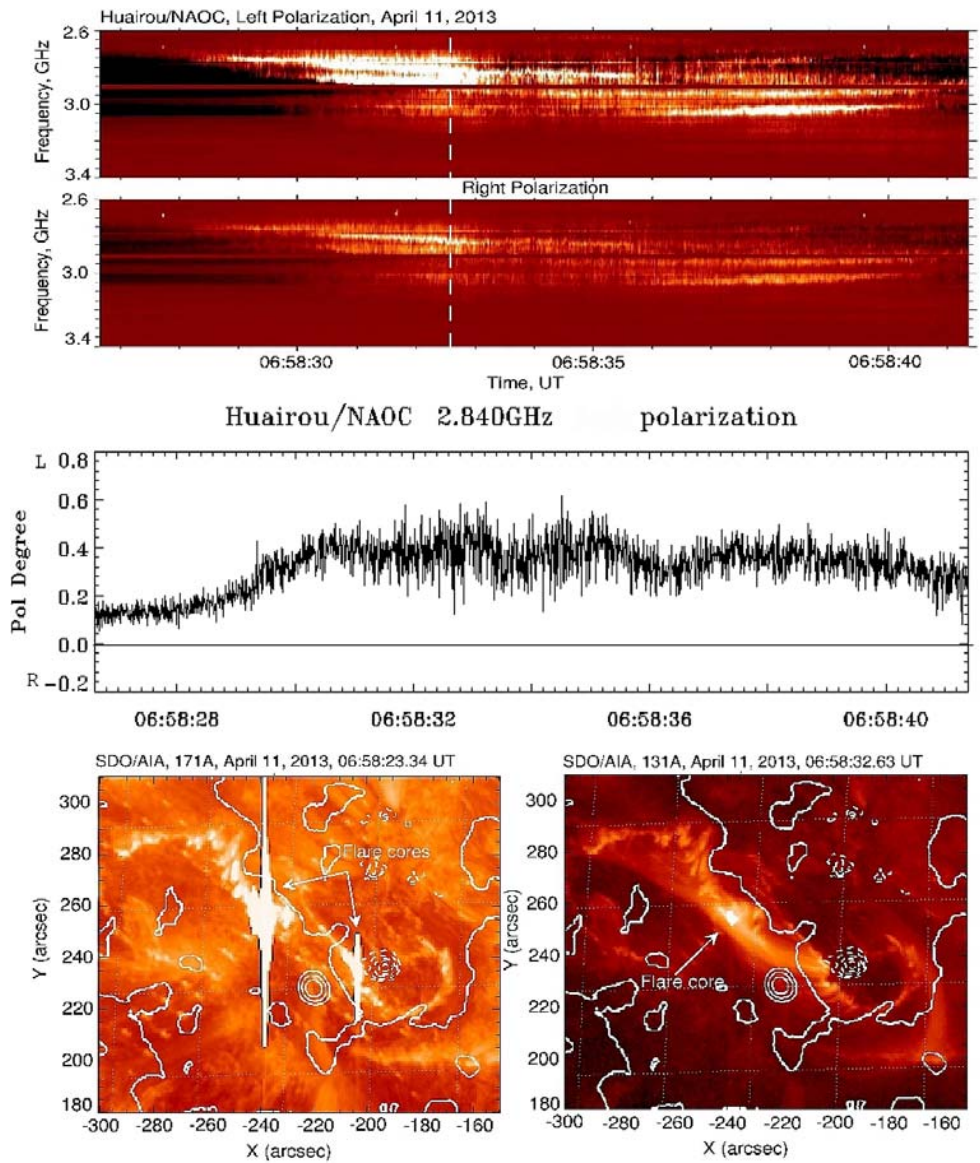

**Figure 15** Strong ZP at the beginning of the event that occurred on 11 April 2013 as registered by the SBRS/Huairou, discussed in Tan et al. (2014). Middle panel: polarization profile at 2.84 GHz. In the bottom: two frames from the movie of SDO/AIA 171 Å (left) and 131 Å (right). The dotted vertical line in the spectrum shows the moment displayed in the right frame at 131Å. Moderate left polarization can be explained by a hard flare core in the left flare ribbon with N magnetic polarity (ordinary wave mode) (Chernov et al. 2016).

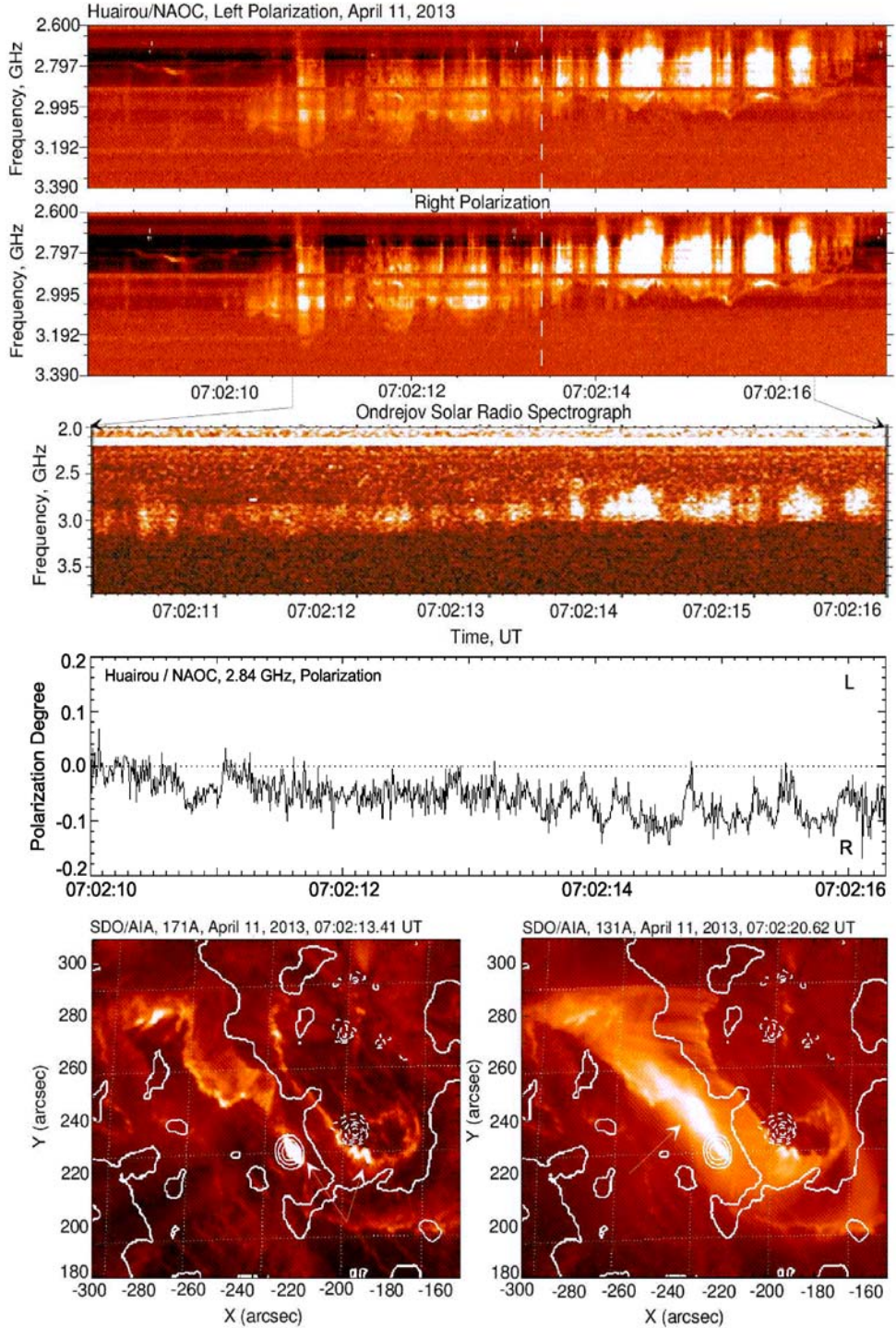

**Figure 16** Fast pulsations during 8 sec registered by the SBRS/Huairou in the 11 April 2013 event. Spectra of pulsations and ZP coincide in observations from the Huairou and Ondřejov observatories. The polarization profile at 2.84 GHz of SBRS/Huairou shows weak right-handed polarization of 8±4 %. Bottom: two frames from the movie of SDO/AIA at 171Å (left) and 131Å (right). The dotted vertical line in the spectrum shows the moment displayed in the left frame in 171Å line (Chernov et al. 2016).

2.2.12. Strange fibers in the event on April 19, 2012

In Figure 17 the unusual type II burst consisting of fibers in the frequency range of 42 – 52 MHz is shown. A stronger emission of type III bursts was superimposed on the fibers. The narrow instantaneous frequency bandwidth of each fiber (</ 1 MHz) allows one to classify them as fiber bursts although they are not parallel to each other.

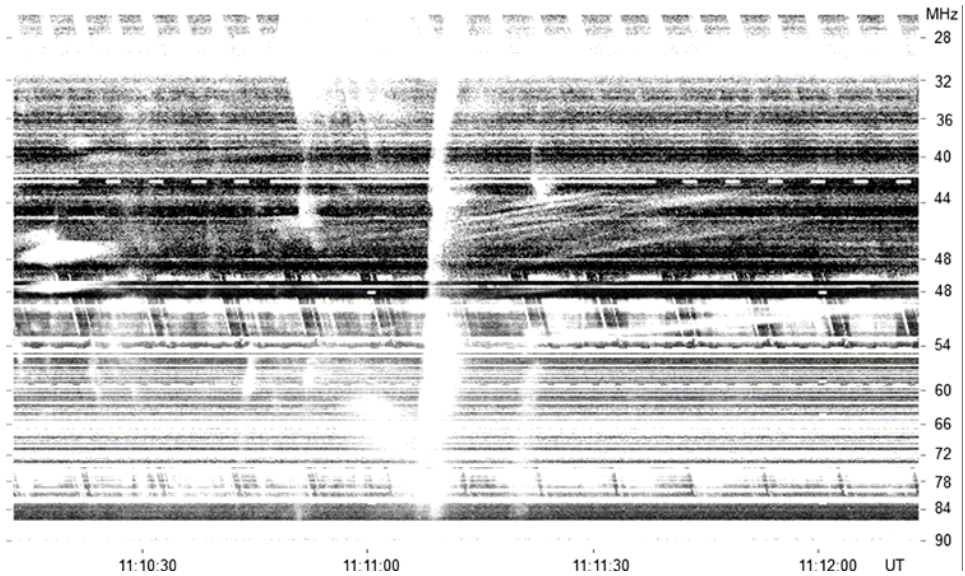

**Figure 17** Type II burst on April 19, 2012 in the range 25 – 90MHz (IZMIRAN) consisting of slow-drifting fibers (from Chernov et al. 2015a).

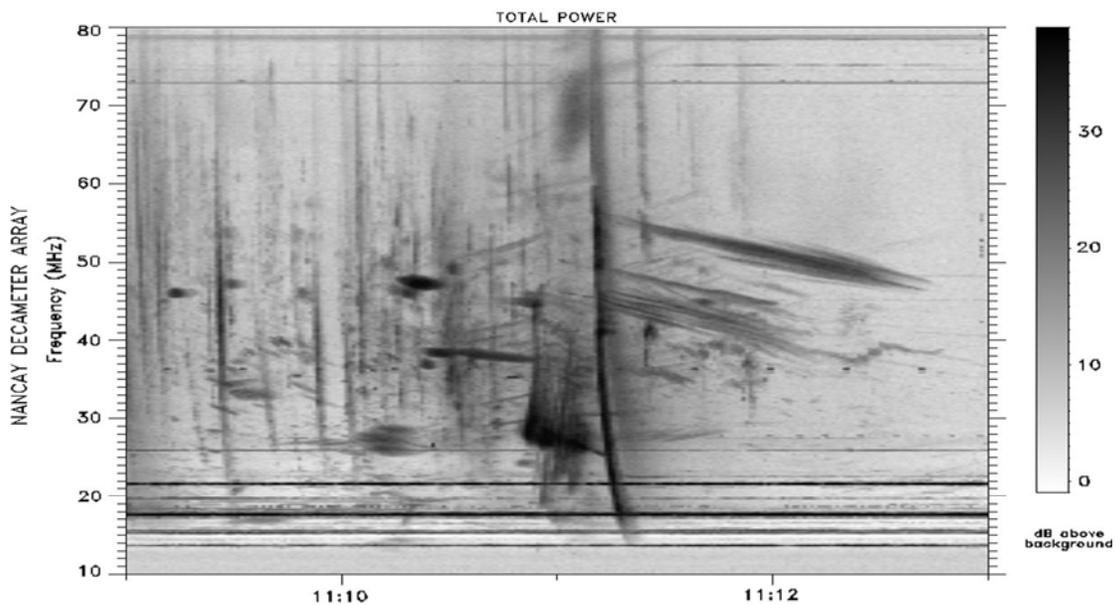

**Figure 18** Type II burst on April 19, 2012 in the range 10 – 70 MHz with Nançay Decametric Array (courtesy Alain Lecacheux).

Figure 18 shows the spectrum of this phenomenon obtained with the Nançay Decametric Array in the range 10 – 80 MHz. All large details match with IZMIRAN spectrum. But here, a more sensitive tool finds other family of weak fibers with a reverse frequency drift and whole ensemble of point-like spikes.

The slow drifting fibers in the type II burst (Figure 17 and 18 ) can be related with standing whistler wave packets before the shock front (Chernov, 1997). A magnetic trap is absent before the shock front, therefore in such a case the quasilinear interaction of whistlers with fast particles is also absent, and low frequency absorption is not formed. Certainly, we are still far from explaining of all components in Figure 18, (besides several families of fibers, numerous point-like spikes). It is possible only to assume the complex turbulent structure before the shock front, or in the intersection region of two shock fronts.

2.2.13.   New instrumentation

Based on an old decimeter solar radio spectrometer working in the frequency range of 625–1500 MHz of the Yunnan Astronomical Observatories (YNAO) during the last solar cycle, Gao et al. (2014) designed a fully digital Fast Fourier Transform (FFT) spectrometer to upgrade the old one. The new digital spectrometer has the spectral resolution of 200 kHz, much higher than the old one (about 1.3 MHz). In addition, they also established a new metric solar radio telescope working in the frequency range of 70–700 MHz located at the Fuxian Solar Observatory of YNAO, deploying the same type of the digital FFT spectrometer. The two instruments have begun to operate in a daily survey mode since September 2009 and March 2012, respectively, and many solar radio bursts have been observed. In these events, various types of decimeter and metric fine structures with fairly meticulous spectral features were recognized. These features were never resolved in previous observation and studies. They have introduced these two instruments with their detailed technological components, as well as a set of observational data obtained during the first-light of the instruments. The information revealed by these data can improve our knowledge and understanding of the physics of the energy conversion, particle acceleration and transportation during the solar eruption.

The following Figures 19, show original fine structures in the August 1, 2010 event, and this confirms rich possibilities of instruments for the future.

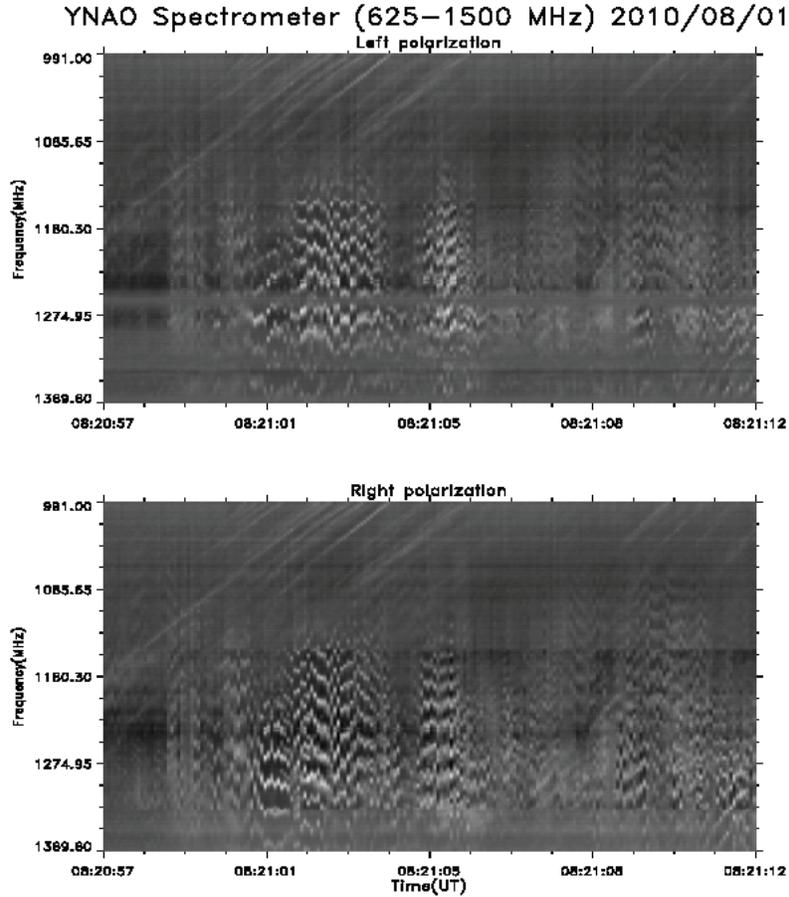

**Figure** 19 Fiber bursts and ZP in decimeter range of the YNAO Spectrometer (private communication from Guannan Gao ggn@ynao.ac.cn ).

Kaneda et al. (2015) investigated the polarization characteristics of a zebra pattern (ZP) in a type-IV solar radio burst observed with AMATERAS on June 21, 2011 for the purpose of evaluating the generation processes of ZPs. Analyzing highly resolved spectral and polarization data revealed the frequency dependence of the degree of circular polarization and the delay between two polarized components for the first time (Figure 20). The degree of circular polarization was 50%–70% right-handed and it varied little as a function of frequency. Cross-correlation analysis determined that the left-handed circularly polarized component was delayed by 50–70 ms relative to the right-handed component over the entire frequency range of the ZP and this delay increased with the frequency. The obtained polarization characteristics were examined by using pre-existing ZP models and concluded that the ZP was generated by the double-plasma-resonance process. The results suggest that the ZP emission was originally generated in a completely polarized state in the O-mode and was partly converted into the X-mode near the source. Subsequently, the difference between the group velocities of the O-mode and X-mode caused the temporal delay.

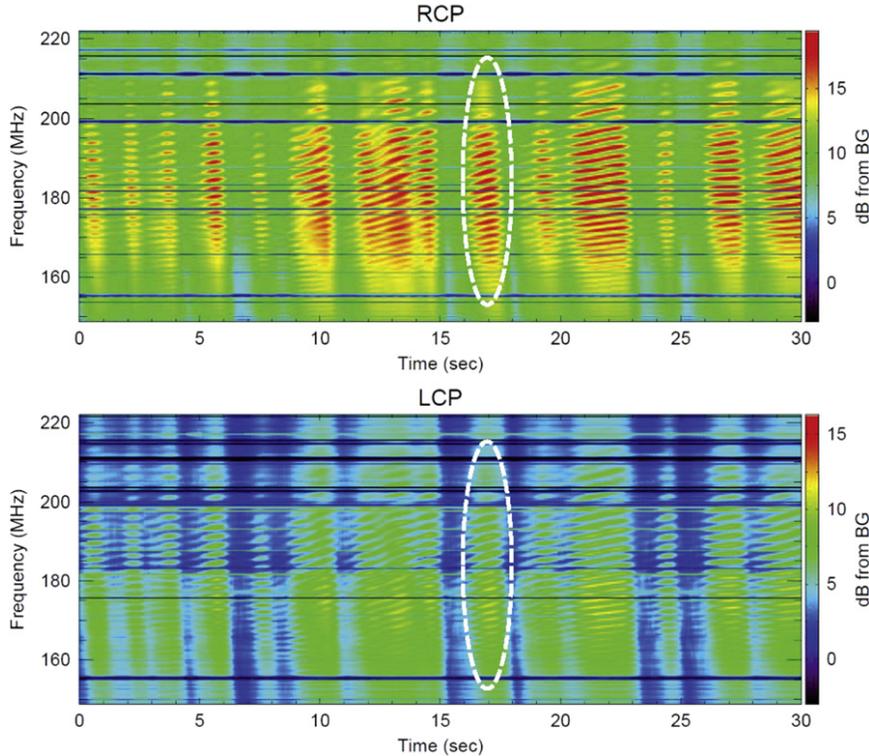

**Figure 20** Magnified image of the ZP for 03:22:06–03:22:36 UT (top: RCP, bottom: LCP). The ZP was enhanced in fast drifting envelopes observed with AMATERAS (Kaneda et al. 2015).

**2.2.14. High resolution observations with Artemis-IV radio spectrograph**
**New paper of Bouratzis et al. (2016) is devoted to observations of spike-bursts in the meter range, and they are of considerable interest as zebra-stripes have superfine structure in the form of ms spikes. In this connection peculiar spikes are of special interest, in particular spike groups exhibiting lace-like morphology, chains of spikes in the form of N burst-like patterns and parallel to fiber bursts (see in Bouratzis et al. (2016) Figures 9, 11, 12).**

2.2.15.
Zlotnik (2013) recalls the main requirements for the instability at DPR. Namely, the background type IV continuum is due to the loss-cone instability of hot non-equilibrium electrons, and the enhanced striped radiation results from the double-plasma-resonance effect in the regions where the plasma frequency $f_p$ coincides with the harmonics of electron gyrofrequency $f_B$, $f_p = sf_B$. Estimations of the electron number density and magnetic field in the coronal magnetic traps, as well as the electron number density and velocities of hot electrons necessary to excite the radiation with the observed fine structure, are given. In particular it is claimed that the wave-like dynamic spectrum of ZP is the radio image of fast magneto-acoustic oscillations of the coronal loop.

2.2.16.
Zlotnik, Zaitsev and Altyntsev (2014) discuss the problem of strong polarization of the zebra-type fine structure in solar radio emission. The degree of polarization of the radio emission at twice the plasma frequency originating from the coalescence of two plasma waves is proportional to the ratio of the electron gyrofrequency to the plasma frequency, which is a small number and is negligible. The outgoing radio emission at the fundamental frequency can be strongly polarized as the ordinary mode due to the escaping conditions. Much attention is given to the examination of depolarization effects in the course of propagation in the corona. It is shown that the depolarization of the radio emission propagating in the corona cannot be associated with the effect of linear coupling of electromagnetic waves in the region of transverse magnetic field.

# 3. ADDITIONAL DISCUSSION

3.1. The study of a short radio burst with rich fine structures on 11 April 2013 showed that each new radio maximum was related to a new flare brightening seen in EUV images of SDO/AIA. Each radio maximum has its own fine structures, usually composed of several stripes of ZP or ZP in the high frequency edge of fast pulsations. Such a relation indicates there is a close connection between radio sources of pulsations and ZP. The movie of SDO/AIA in 131 Å showed a flare loop arcade forming between two sigmoid flare ribbons. Therefore the flare dynamics consisted of consecutive magnetic reconnections in different arcade loops.

The polarization changed in accordance with the position of the new flare brightening. The left flare ribbon was located above the North magnetic polarity (tail spot) and the right ribbon above the South magnetic polarity (leading spot). In all cases, the radio emission mode remained ordinary. When the brightening took place at the looptops, the polarization was very weak, almost zero.

The magnetic field remained stable during the event. At the same time it was improbable that motion of the radio source from one flare ribbon to another one lasted several seconds. A similar explanation of a gradual changing of the polarization sign at 17 GHz (Nobeyama data) was proposed by Huang and Lin, 2006.

Only one question arises: why we receive only a partial degree of polarization? If the emission is generated at the fundamental (by some mechanism) as the ordinary mode, it is fully polarized in the source. During propagation of the radio waves the observed polarization degree is changed due to a depolarization effect. The depolarization happens in a layer where the radio emission is propagated exactly across the magnetic field. In the considered event, the source geometry (the flare occurred at the disk centre) allows that this condition is possibly satisfied for a ZP source in a closed magnetic trap (for more details see Chernov, Zlobec (1995)).

The emission of fast radio pulsations is probably caused by fast electrons accelerated in the upward direction in a vertical current sheet with the same period during magnetic reconnection. Some of the fast particles that are accelerated downward can be captured in a closed magnetic trap and they could be responsible for the emission of ZP by some single mechanism. The diversity of ZP stripes is probably given by different conditions in different arcade loops. This is natural scheme of flare processes which is in accordance with the standard model of the solar flare as shown in Figure 19.

Such an expected position of a microwave ZP source at tops of flare magnetic loops was also confirmed in a new paper by Yasnov and Karlický (2015) in results of estimation of bremsstrahlung and cyclotron absorptions of radiation in the corona.

We do not have any convincing evidence that the mechanism that generated the ZP is the emission of Bernstein modes as this was proposed by Tan et al. (2014b) for the first strong ZP. First, an exact definition of the frequency separation between three (and sometimes four) zebra stripes is too problematic. Second, the polarization should be related to the extraordinary emission mode (Zlotnik, 1976; Kuznetsov, 2005). Third, we do not have any information about size of the radio source with height in the corona (distributed or point like). Else, the radio emission defined by Bernstein modes must be weak, much weaker than in other mechanism, i.e., in the double plasma resonance (Zlotnik, 2009) or interaction of Langmuir waves with whistlers (Chernov, 2006). In both of the last models, the radio source should be distributed in heights, but in the double plasma resonance model a source should be stationary and in the whistler model – moving (depending on the group velocity of the whistler wave) (for more details see Chernov, Yan and Fu (2014). Furthermore, the spatial drift of ZP stripes should change synchronously with changes in the frequency drift of the dynamical spectrum. Recently, more than ten other mechanisms were proposed for ZP, but their significance remains uncertain (Chernov et al. 2014).

To choose which mechanism applies, positional observations may be crucial, and it is desirable to observe a limb event. Now, we are expecting progress in the field of solar radio imaging spectroscopy. The first trial observations began on the new Chinese spectral radioheliograph (CSRH) which will be the largest and most advanced radio imaging telescope for the solar corona in the world (Yan et al. 2009; Yan et al. 2013).

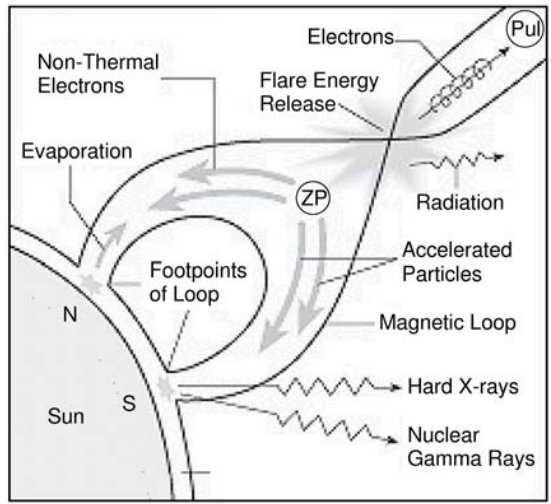

**Figure 21** Expected radio source positions of ZP and pulsations (Pul) in the scheme of the standard flare model of Aschwanden, 2006.

In several events we have opposite combinations of a ZP and fiber busts, when fibers limit the ZP emission to the high frequency part in the decimeter range in the 1 December 2004 event (see Figure 4 in Chernov et al. 2014) and microwave range in the 1 August 2010 event (see Figure 3 in Chernov et al. 2014). For such events we propose another combinations of radio sources (Figure 20 ) when ZP is emitted from a magnetic island.

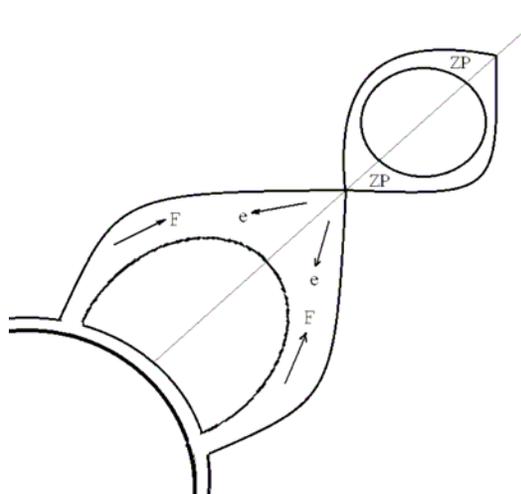

**Figure 22** Expected radio source positions of ZP and fiber bursts in events at 1.12.2004 and 1.08.2010.

3.2. Many debatable moments were considered in Chernov et all. (2015a).

The new examined events show that the zebra-structure and fiber bursts can appear almost simultaneously or consecutively in the microwave, decimeter and meter wave bands. In the December 1, 2004 and August 1, 2010 events they were almost superimposed on each other. Certainly this fact was known earlier. For example, in the meter range see Fig. 4.9 in Chernov, 2011 and in the microwave – ibid, Fig. 4.38. However when interpreting such phenomena authors usually artificially mark out two idealized structures: parallel drifting (probably wavy) stripes of the zebra- structure and fiber bursts with intermediate frequency drift (constant and negative). Though the phenomena in which stripes of the zebra structure smoothly pass to fibers and back are known (see, for example, tin monographs Chernov (2011) fig. 4.13 and 4.22). In the event on 15 February 2011 we also have an example (Figure 8) of a such passage of one structure into another (according to their determination).

The weak zebra (Fig. 2a and Fig. 8 in Chernov et al. 2015a) can be explained simply in the whistler model, when periodic whistler wave packets fill in the whole magnetic trap. The appearance of fiber bursts in the same frequency band where two minutes before fast pulsations were observed (Figure 7), tests if radio source for both structures are the same, and both mechanisms are related with whistlers (see Ch.1 in Chernov (2011).

In the theory of formation of these structures most often any smooth transition of one structure to another is not considered at all. In the theory of zebra- structure the wide acceptance obtained the mechanism of double plasma resonance (DPR), when the upper hybrid frequency becomes equal to an integer of electronic cyclotron harmonics. Fiber bursts are explained by a completely different mechanism, the interaction of plasma waves with the whistlers, propagating with the group velocity in the form of the periodic wave packets. The DPR- mechanism obtained wide recognition due to the very detailed developments of Zheleznyakov (1995) and subsequent series of articles and reviews of Zlotnik (Zlotnik et al. (2003). Zlotnik (2009), Zlotnik, (2010)).

Really, the theory archives completeness of analytical analysis as well as the source model by its simplicity. The impression of the authenticity of model and operation of the mechanism is immediately created. This allows us to assume that the DPR- mechanism must always work, if there is a magnetic trap, in which the plasma density and the magnetic field strength decrease smoothly with the height in the corona with different gradients. A question simultaneously arises, why then does zebra-structure only appears irregularly (by per-second intervals) in the prolonged phenomena. In the images of SOHO/EIT then of TRACE and now also of SDO/AIA the magnetic flare loops remain almost constant during the entire phenomenon. The repetitive maximums of continuous emission confirm the presence of fast particles. However, zebra-structure only appears irregularly during several seconds.

Up to now, we do not have any information about the operation of the DPR mechanism at several resonance levels in laboratory plasma experiments (see also below 4.5) although the simulation of magnetic reconnection and excitation of ion-acoustic and whistler waves was realized long ago. The observations of the last few years confirm the superfine structure of flare magnetic loops. In a number of works Aschwanden has shown that in the thin loops the magnetic field weakly changes with height (Aschwanden, 2004). In such loops it is not possible to expect even two DPR- levels, even with the most varied assumptions about the decrease in the density with height. It is possible to recall that in some phenomena more than 30 zebra stripes were observed simultaneously in the range 2.6 - 3.8 GHz. Thus, according to Fig. 5.21 in the book Chernov (2011) it is shown that in any known models of coronal plasma it is impossible to obtain the large number of DPR-levels.

Nevertheless no authors consider it important to demonstrate the possibility of existence of the DPR levels in space plasma. Moreover, scale heights are taken arbitrarily in order to obtain several intersections of the curves of plasma frequency and electronic cyclotron harmonics as, for example this is done in the recent paper Chen et al (2011). It is considered that the DPR- mechanism explains any phenomena with a zebra structure. In practically all works with the discussion of DPR- mechanism the presence of the large number of DPR- levels is considered obvious. Only in Zlotnik et al (2003) is an attempt undertaken at the more concrete specific selection of magnetic trap with the aid of the calculated magnetic map above AR for the phenomenon on October 25, 1994. However, the selection of a basic loop was made arbitrarily (more precise erroneously) without taking into account the fact that the particles critical for III type bursts and zebra- structure moved in the different directions (for greater detail, see Chernov et al (2005)).

In several resent works the authors began to compare DPR- models and the interaction of plasma waves with whistlers. In this case an inaccuracy in the estimations of different parameters of whistlers were allowed. In this connection, it is necessary to note that the model with whistlers was mistakenly rejected in Chen et al (2011) (see above). In Tan et al. (2012) the estimation of the magnetic field strength using the frequency drift of fibers in the model with the whistlers is made for the frequency of whistlers relative to electron cyclotron frequency $x = f_w/f_{ce} = 0.01$, and during the estimation of field according to the frequency separation of zebra strips in the same frequency range it is taken as $x = 0.25$. As a result in the latter case the field strength is underestimated by twice, and model with the whistlers is rejected. The value $x$ is determined by the frequency, at which the value of the increment of whistlers is maximum. According to calculations of Chernov (2011) the value $x$ is located in the interval of 0.1 to 0.01 depending on different parameters of the distribution function of fast particles. Therefore the values of field strength in the model with the whistlers must exceed estimations obtained in the DPR-model.

In the critical review of Zlotnik (2009), the advantages of the DPR model and the main failures of the model with whistlers are refined. The author asserts that the theory based on the DPR effect is the best-developed theory for the origin of ZP at meter-decimeter wavelengths at the present time. It explains the fundamental ZP feature in a natural ways, namely, the harmonic structure (frequency spacing, numerous stripes, frequency drift, etc.) and gives a good fit for the observed peculiarities in the radio spectrum with quite reasonable parameters of the radiating electrons and coronal plasma. The statement that the theory based on whistlers is able to only explain a single stripe (e.g., a fiber burst) was made in Zlotnik (2009) without the correct ideas of whistler excitation and propagation in the solar corona.

Zlotnik uses the term "oscillation period" of whistlers connected with bounce motion of fast particles in the magnetic trap. Actually, the loss-cone particle distribution is formed as a result of several passages of the particles in the magnetic trap. Kuijpers (1975) explain the periodicity of the fiber burst using this bounce period (~1 s). If we have one fast injection of fast particles, whistlers (excited at normal cyclotron resonance) are propagated towards the particles (they disperse in the space). Quasilinear effects thereby do not operate in normal resonance.

A ZP is rather connected with whislers excited at anomalous resonance during long lasting injection. In such a case, waves and particles propagates in one direction, quasilinear effects begin operate and their role increases with increasing duration of injections. A ZP is excited because the magnetic trap should be divided into zones of maximum amplification of whistlers, separated by intervals of whistler absorption (see more details in Chernov (1990)). The bounce period does not interfere with this process, but it can be superimposed on the ZP.

However, the whistler amplification length is always small (on the order of $\leq 10^8$ cm in comparison with the length of the magnetic trap being $>10^9$ cm) for any energy of fast particles (Breizman, 1987, Stepanov and Tsap, 1999). According to Gladd (1983), the growth rate of whistlers for relativistic energies of fast particles decreases slightly if the full relativistic dispersion is used. In this case, the whistlers are excited by anisotropic electron distributions due to anomalous Doppler cyclotron resonance.

Later, Tsang (1984) specified calculations of relativistic growth rates of whistlers with the loss-cone distribution function. It was shown that relativistic effects slightly reduce growth rates. According to Fig. 8 in Tsang (1984), the relativistic growth rate is roughly five times smaller than the nonrelativistic growth rate. However, the relativistic growth rates increase with the perpedicular temperature of hot electrons $T$. According to Fig. 5 in Tsang (1984), the growth rate increases about two times with increases the electron energy from 100 to 350 keV, if keeping other parameters of hot electrons: loss-cone angle, ratio of gyrofrequency to plasma frequency, temperature anisotropy ($T/T_\parallel = 3$).

Thus, it was known tlong ago hat the whistlers can be excited by relativistic beam with loss-cone anisotropy. Formula 13.4 in Breizman (1987), used as formula (29) in Chernov 2006) for evaluating the smallest possible relaxation length of beam, has no limitations in the value of energy of fast particles.

A critical comparison of models has been repeated in Zlotnik (2010), only with a new remark concerning the Manley-Rowe relation for the brightness temperature of electromagnetic radiation that is a result of coupling of Langmuir and whistler waves:

$$T_b = \frac{\omega\, T_l T_w}{\omega_l T_w + \omega_w T_l}. \qquad (2)$$

Zlotnik (2010) states that since $\omega_w \ll \omega_l$, in the denominator, only the first term remains and $T_b$ depends only on $T_l$, and $T_b \sim T_l$, i.e. the process does not depend on the level of whistler energy. However, Kuijpers (1975) (formula (32) in page 66) shown that the second term $\omega_w T_l$ should be $\gg \omega_l T_w$ due to $T_l \gg T_w$. An analogous conclusion was made by Fomichev and Fainshtein (1988) with more exact relation with three wave intensities (see also formula (11) in Chernov (2006)). Therefore the value of $T_b$ in the process $l + w \rightarrow t$ depends mainly on $T_w$. Thus, our conclusion, that the entire magnetic trap can be divided into intermittent layers of whistler amplification and absorption remains valid for a broad energy range of fast particles.

In Zlotnik (2009) the main matter which is ignored is that the model involves quasilinear interactions of whistlers with fast particles, allowing one to explain all the fine effects of the ZP dynamics, mainly the superfine structure of ZP stripes and the oscillating frequency drift of the stripes which occurs synchronously with the spatial drift of radio sources. For an explanation of the superfine structure the alternative whistler model remains as the most natural: dynamic energy transfer between ion-sound and whistler waves in a pulsating regime with the process $s + s' \rightarrow w$ (Chernov, Yan, Fu, 2003; Chernov, 2011, section 4.7.5).

Similar continuous discussions stimulate the developments of new models. Treumann et al. (2011) proposed new mechanism of ZP, the ion-cyclotron maser. Thanks to the special delta-shaped distribution function of the accelerated ions, the ion-cyclotron maser generates a number of electromagnetic ion-cyclotron harmonics which modulate the electron maser emission. A part of the accelerated relativistic protons passes along the magnetic field across the trapped loss-cone electron distribution. The modulation of the loss-cone will necessarily cause a modulation of the electron cyclotron maser. Locally this produces the typical "Zebra" emission/absorption bands. However this mechanism can work in the strong magnetic field, when $f_{pe}/f_{ce} < 1$.

**4. RECENT RESULTS ON THE IMPROVEMENT OF THE DPR MODEL**

Karlicky made a big contribution to investigate or improve the DPR mechanism.

4.1. Karlický et al. (2013) continued the development of the model of Kuznetsov (2006) for fiber bursts. Kuznetsov (2006) proposed a model in which the fiber bursts are generated by a modulation of the radio emission by magneto-hydrodynamic waves. He also proposed that these waves could be magneto-acoustic waves of a sausage mode type that propagate along a dense coronal loop. Fiber bursts can be explained by the propagating fast sausage magneto-acoustic wave train. Then Karlický (2013) extended a similar model for ZP: the magneto-acoustic waves with density variations modulate the radio continua, and this modulation generates zebra effects. It should be noted that close model (with density heterogeneities) was examined in works by Laptuhov and Chernov (2009; 2012).

4.2. Karlický and Yasnov (2015) present a new method of determination of the magnetic field strength and plasma density in the solar zebra radio sources.

Using the double plasma resonance (DPR) model of the zebra emission they derived analytically the equations for computations of the gyroharmonic number $s$ of selected zebra line and then they solved these equations numerically. The method was successfully tested on artificially generated zebras and then applied to observed ones. The magnetic field strength and plasma density in the radio sources were determined. Simultaneously, they evaluated the parameter $L_{nb} = 2L_b/(2L_n - L_b)$, where $L_n$ and $L_b$ are the characteristic scale-heights of the plasma density and magnetic field strength in the zebra source, respectively. Computations show that the maximum frequency of the low-polarized zebras is about 8 GHz, in very good agreement with observations. For the high-polarized zebras this limit is about 4 times lower. It was shown that microwave zebras are preferentially generated in the regions with steep gradients of the plasma density as for example in the transition region, on the basis of the previous results of the authors (Yasnov and Karlický, 2015) using density model by Selhorst, Silva-V´alio, and Costa (2008) (Figure 23). In models with smaller density gradients as e.g. in those with the barometric density profile, the microwave zebras cannot be produced owing to the strong bremsstrahlung and cyclotron absorptions. They also showed that this DPR model is able to explain the zebras with frequency-equidistant zebra lines.

The bremsstrahlung absorption in atmospheric layers above the DPR zebra generation region and the cyclotron absorption in the DPR region and in the gyroresonance layers at higher altitudes limit the spectrum of zebras and micro bursts (MB) from both high-frequency and low-frequency sides.

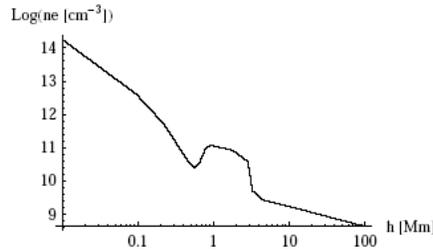

**Figure 23** The electron density profile as a function of height in the solar atmosphere according to Selhorst et al. (2008).

4.3. Karlický (2013) extended his model to zebra discovered in radio emission of Crab pulsar: Zebra patterns similar to those analyzed in solar observations were observed in the radio emission of the Crab Nebula pulsar (Hankins & Eilek 2007), see Figure 24.

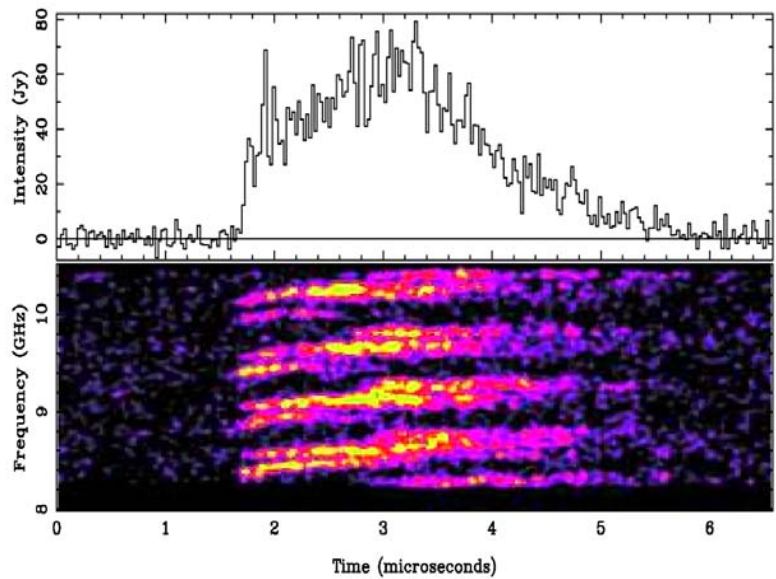

**Figure 24** Zebra pattern spectrum in microwave emission from the Crab Nebula pulsar (Hankins & Eilek 2007).

Recently, these zebras were interpreted by Zheleznyakov et al. (2012) using the double plasma resonance model. This explains many observed features of these zebras. However, the main problem of the model is that it requires relatively low magnetic field in the pulsar radio source.

Simulations of Karlický (2013) assume that in the pulsar atmosphere the radio continuum is generated at the double-plasma frequency $2\omega_{pe}$ from two coalescing plasma (Langmuir) waves. Furthermore, in the radio continuum source they assumed a fast magneto-acoustic wave. This wave propagates at Alfvén speed, which for our case with a very high magnetic field $B$, is close to the speed of light $c$.

Karlický proposes that the frequency drift of the observed zebra is caused by the propagating fast magneto-acoustic wave. It gives us the spatial scale in the density model. The wave velocity is $v_A \approx c$, the used relative amplitude of density wave and the wave wavelength are $n_R = 0.01$ and 1.5 km. Due to the positive frequency drift of the observed zebra lines, the assumed wave propagates toward lower heights in the present density model. He used the same procedure as in the solar case, he received a computed radio spectrum. To better specify them a more specific model of the pulsar radio continuum is needed.

In other models, e.g., in the model of Hankins & Eilek (2007), the radio source is in the highly relativistic jet.

It should be noted that no one takes into consideration the basic special features of the pulsar zebra: stripes have dual structure, four large envelopes are splitting into 2-3 the narrow stripes, and the duration (microseconds) of 6 orders less than the solar one.

4.4. Fingerprint ZP

At last, in the recent paper of Zlotnik, Zaitsev, Melnik et al. (2015) a peculiar fine structure in the dynamic spectrum of the solar radio emission discovered by the Radio Telescope UTR-2 spectrograph (Kharkiv, Ukraine) in the frequency band 20-30MHz is discussed. The structure is observed against the background of a broadband type IV radio burst and consists of the parallel drifting narrow bands of enhanced (versus the background) emission and absorption (Figure 25). The observed structure differs from the widely known zebra pattern at the meter and decimeter wavelengths by the opposite directions of the frequency drift within the limits of a single stripe at a given time. It is shown that the observed peculiarities can be understood in the framework of the plasma mechanism of the radiation origin by virtue of the double plasma resonance effect in a nonuniform coronal magnetic trap. The source model providing the peculiar frequency drift of the zebra stripes is proposed.

The authors thoroughly selected the numbers of harmonics and showed that the best fit with the observed forms of stripes ("fingerprint" or vertical arcs) gives the collection of the harmonics $s =20$-30. They used the slightly modified Newkirk model for the density and the specific functional form of the magnetic field on height ($B = B_0(t)/(1 +h/7.95)^3$). The magnetic field gradient must be very similar to the electron number density gradient with opposite signs of these gradients at the upper and lower parts of the source (to provide different signs of the frequency drift in the high- and low-frequency parts of the spectrum). At each time no more than 1-3 harmonics are observed, and each harmonic must twice intersect the curve of plasma frequency in each moment (Figure 26 ). Authors conclude, that the presented model of the source is able to explain the details of the unusual frequency drift of the stripes of enhanced intensity in the observed fingerprint spectrum within the framework of the DPR mechanism under quite reasonable physical conditions in the coronal magnetic trap.

In this connection, it worth to note that intensity time profiles at frequency 25.45 MHz shown in Figure 27 (from previous paper of Melnik et al. 2008) presents more evident absorptions of the continuum level. Besides, the differential spectrum at the bottom of Figure 25 show evident disruptions of stripes at frequency 25.5 MHz (just a nose frequency). Other elements of the continuum emission don't show any disruptions.

The solar origin of this unusual ZP we could confirm by IZMIRAN spectrum (Figure 28) where some zebra-stripes are visible in the designated oval, although the instrument had of one order smaller sensitivity, and it was in average in time regime by 1 s.

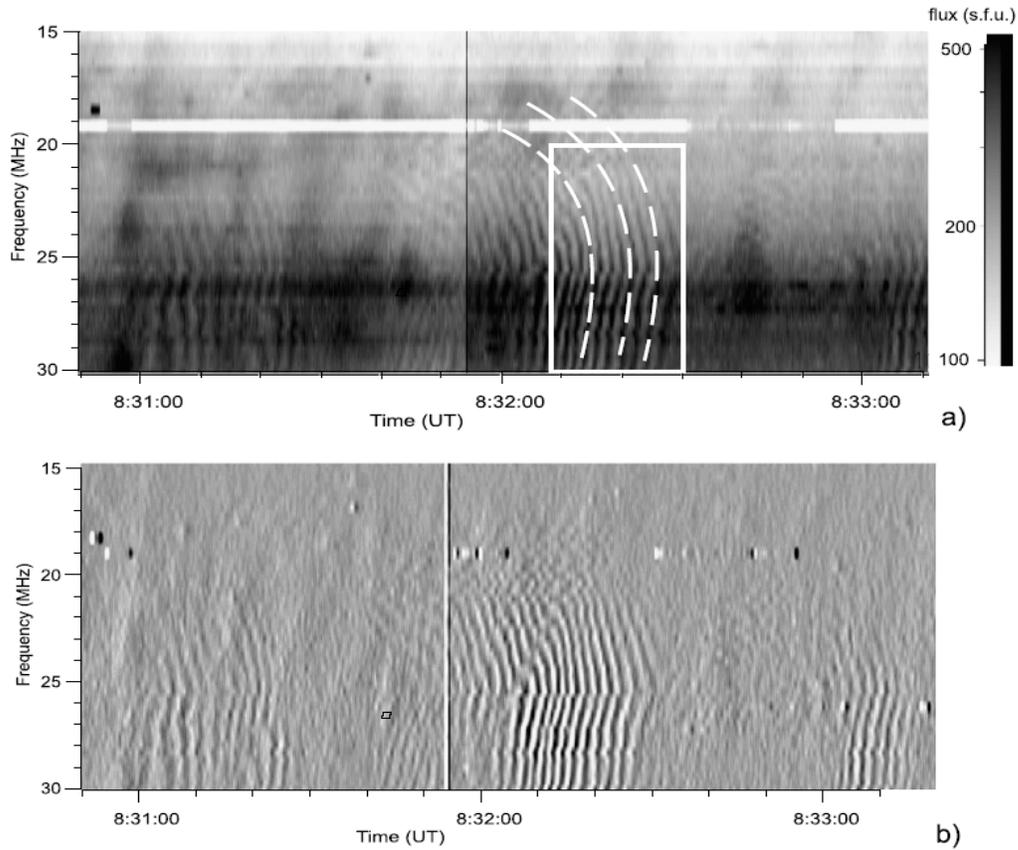

**Figure 25** Dynamic spectrum of the solar burst 'fingerprint' on 22 July 2004 (a) and its differential dynamic spectrum (b). The rectangle indicates the portion that is used for the numerical analysis. The dashed lines schematically show the form of the fine structure line.

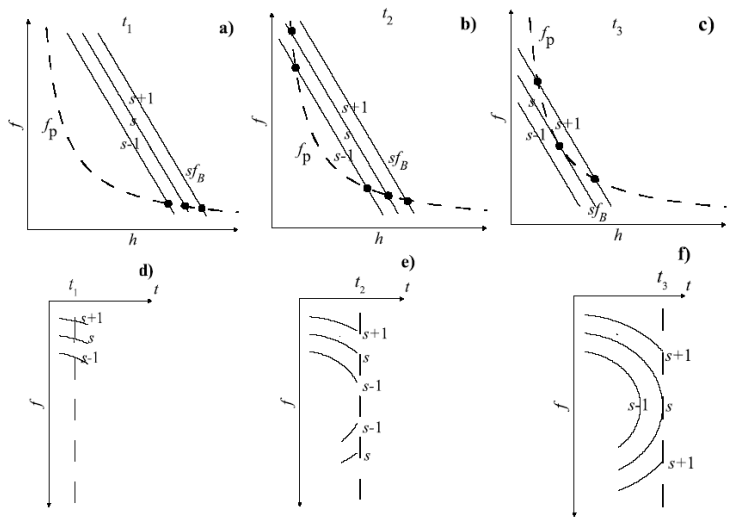

**Figure 26** Qualitative interpretation of the fingerprint structure as a set of quasi-vertical arcs. The top panel is the source model at three times $t_1 < t_2 < t_3$; the electron number density remains constant, and the magnetic field decreases in time. The filled circles denote the DPR levels. The lower panel gives the dynamic spectrum with the zebra stripes corresponding to the times *t*1, *t*2, and *t*3.

It is possible to still note (as the resume), the decrease of magnetic field utilized for explaining the frequency drift by fast magneto-acoustic wave (FMA) seems improbable at the heights in the corona of order of one solar radius. The life time of the zebra is evaluated by FMA wave period, i.e., without the FMA wave the DPR mechanism does not work? And really, Karlický (2013) extended a model of Kuznetsov (2006) for fiber bursts, for ZP: the magnetoacoustic waves with density variations modulate the radio continua, and this modulation generates zebra effects.

Disruption of the zebra stripes just at nose frequencies says that stripes drifted from different levels independently.

In such a case, the model of the ZP could be much simpler. The ZP appeared simultaneously with coronal mass ejection after which a magnetic islands were formed wherein the fast particles were accelerated simultaneously in top and bottom X- points of the magnetic reconnection. Periodical whistler wave packets generated by these particles propagated towards each other and their interaction with plasma waves gives stripes in emission and absorption. The group velocity of whistlers for the ratio $f_p/f_b$ =20 will be about $10^9$ cm/s.

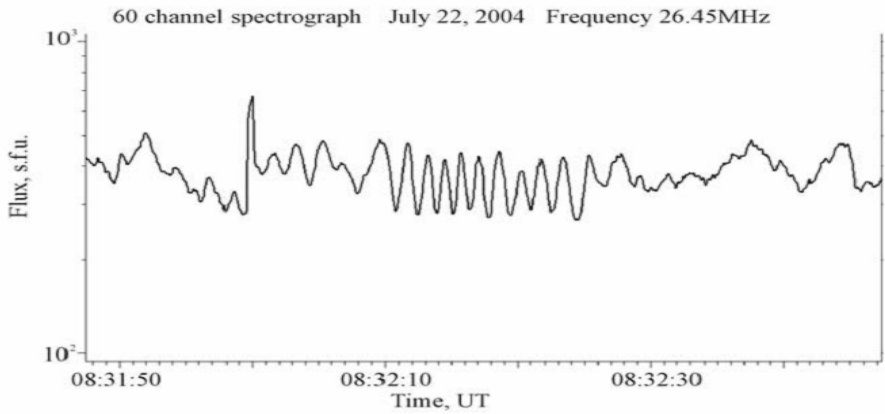

**Figure 27** Intensity profile at 25.45 MHz of ZP in the July 22, 2004 event observed by spectrograph of the Radio Telescope UTR-2 (from Melnik, Rucker and Konovalenko, 2008).

The estimation of possible positive frequency drift (for instance, between 20 and 25 MHz) $df/dt = -f V_{gr}/2L_n$ where $L_n$ is the scale height of density of = $10^{10}$ cm gives about the same value as observed, of 1MHz/s.

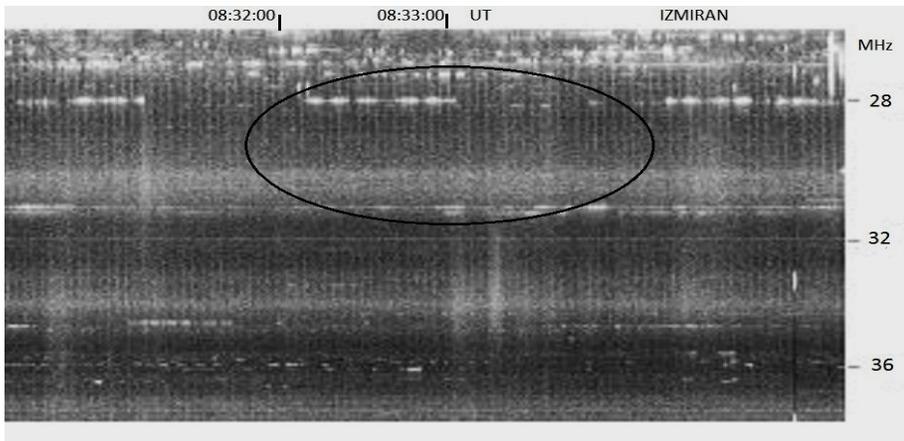

**Figure 28** IZMIRAN spectrum of the July 22, 2004 event in the range 25-40 MHz.

The radio source of the ZP was evidently located in the tail of the CME, in a inhomogeneous plasma, therefore it is possible to examine also the probability of other mechanisms, for example radio wave propagation through small heterogeneities in the tail of CME (Laptuhov and Chernov, 2009). However, in this case two sources must also emit simultaneously (above and below magnetic cloud), and heterogeneities must drift towards each other. Laptuhov and Chernov (2012) showed that alternating transparency and opacity stripes in the spectrum of radio waves passing through such a plasma structure (the zebra pattern effect) can be observed at any angle of incidence.

4.5. **Laboratory plasma experiments**
We found only one confirmation about the operation of DPR- mechanism in the laboratory installations but only at the second cyclotron harmonic (Figure 29 ) in Viktorov, Mansfeld and Golubev (2015a). In this paper they report about kinetic instabilities of nonequilibrium plasma heated by powerful radiation of the gyrotron under the electron cyclotron resonance (ECR) conditions and confined in the mirror magnetic trap. It is seen in Figure 29 that any emission at third cyclotron harmonic is absent. It is possible certainly to connect this with the difficulties of designing in the installation of the parameters of plasma, similar in the magnetic trap in the sun, mainly the ratio $f_p/f_b >> 1$.

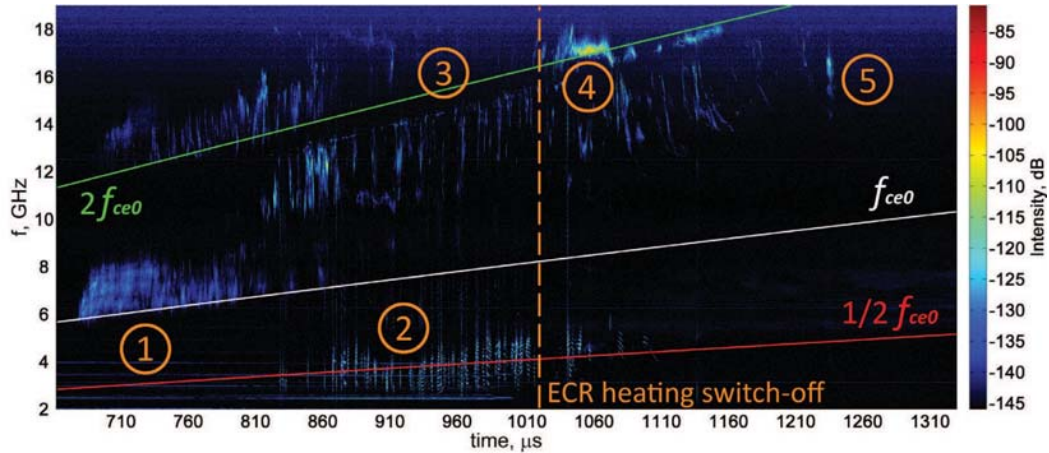

**Figure 29** Dynamic spectrum of the plasma radiation. The types of kinetic instabilities considered in the paper are highlighted in the figure: 1: the initial stage of the ECR discharge (rarefied plasma); 2 and 3: stages of the developed discharge (dense plasma); 4: the beginning phase of plasma decay (dense plasma); 5: decaying plasma (rarefied plasma). The three lines on the spectrogram show the time variation of the corresponding frequencies $2f_{ce0}$ (green), $f_{ce0}$ (white) and $1/2f_{ce0}$ (red), where $f_{ce0}$ = $f_{ce}$ (z center) is the electron cyclotron frequency in the center of the magnetic trap on its axis (a part of Fig. 2 from Viktorov, Mansfeld and Golubev (2015a).

Emission of dense plasma at frequencies about $f_{ce0}/2$ (type 2 in fig. 29) is more likely related to the whistler mode instability. The observed radiation at higher frequencies with the sharp upper spectral boundary (type 3) apparently can be related to the excitation of plasma waves under upper hybrid resonance.

Similar experiment is discussed also in Viktorov, Golubev, Zaitsev and Mansfeld (2015b), where the DPR is discussed in more details. The main attention is devoted to fast temporal structure of observed radio emission.

## 5. Conclusions

We have shown that the polarization of the ZP corresponds to the ordinary wave mode and it changes in accordance with dynamics of flare processes. Simultaneous or consecutive appearance of zebra-structure in different frequency ranges is obviously connected with the dynamics of flare processes.

A future analysis needs to clarify whether a radio source showing ZP is really related to a closed magnetic loop, and if it is located at lower altitudes than the source of the pulsations, as expressed on the radio spectrum by ZP at the high frequency boundary of pulsations. New solar radio spectral imaging observations should help to compare the source sizes of different fine structures, and the main thing to determine whether the radio source does move.

In the whistler model radio sources of fiber bursts and ZP are moving, and the spatial drift of ZP stripes should change synchronously with changes of the frequency drift in the dynamical spectrum. In the model of double plasma resonance, the ZP source must be rather stationary.

In almost all papers the interpretation of ZP begins with the most accepted DPR model. Here, we pay attention to several problems with this model in the explanation of complicated new events. Thus, we showed that the event from December 14, 2006 examined in the highly cited paper of Chen, Bastian et al. (2011) in DPR model, could be likewise explained in the whistler model, taking into account the propagation of whistlers along a magnetic trap.

The "fingerprint" form of the ZP in the decameter range is also discussed in the DPR model, whereas it can be simply related with whistler propagation in magnetic islands. Zlotnik et al. (2015) explain the fast frequency drift and lifetime of the ZP with fast magneto-acoustic waves which modulate the DPR mechanism, while Karlický (2013) used FMA density waves (modulating the continuum emission) as a source of ZP without any attraction of the DPR. At the same time Yu, Nakariakov et al. (2013) by FMA oscillations, explain only the quasi-periodic wiggles of a microwave ZP.

The DPR mechanism was even used to interpret zebra-like stripes in microwave radiation of the Crab Nebula pulsar, despite numerous problems with plasma parameters, and Karlický, instead of this, proposed FMA waves once more.

Simultaneous appearances of fibers and zebra-structure and the smooth transition of zebra-stripes into the fibers and back verify the united mechanism of the formation of different drifting stripes in emission and absorption within the framework of the interaction of plasma waves with whistlers, taking into account quasi-linear interaction of whistlers with the fast particles and with ion-acoustic waves.

The model with whistlers explains many special features of the zebra- structure:
- weak zebra stripes with different scales in the broadband frequency range;
- simultaneous appearance of fiber bursts, fast pulsations and spike-bursts;
- the oscillatory frequency drift and the frequency splitting of stripes;
- a change in the spatial drift of radio source synchronously with the frequency drift of stripes in the spectrum;
- the millisecond superfine structure of stripes.
- the rope-like fibers.

The presence of ion-acoustic waves can be considered justified in the source in the region of magnetic reconnection with the outgoing shock fronts. The propagated ion-acoustic waves can serve as natural heterogeneities, passage through which electromagnetic waves can lead to shaping the additional stripes of transparency and opacity on the spectrum. Laptukhov and Chernov ( 2012) showed that alternating transparency and opacity stripes in the spectrum of radio waves passing through such a plasma structure (the zebra pattern effect) can be observed at any angle of incidence. It should be noted that the relative significance of several recent possible mechanisms remains uncertain.

Simultaneous or consecutive appearance the zebra- structure in different frequency ranges is obviously connected with the dynamics of flare processes.

For a comparative analysis of observations of zebra structure and fiber bursts and different theoretical models with many debatable moments we refer the reader to the review by Chernov (2012).

**Acknowledgements**

The author are grateful to the RHESSI, SOHO (LASCO/EIT), SDO and Nobeyama Radioheliograph teams for operating the instruments and performing the basic data reduction, and especially, for the open data policy. The research that was carried out by G.P. Chernov at National Astronomical Observatories (NAOC) was supported by the Chinese Academy of Sciences Visiting Professorship for Senior International Scientists, grant No. 2011T1J20. This work is also supported by the Russian Foundation for Basic Research under Grant 14-02-00367.

**References**

Altyntsev, A.T., Lesovoi, S.V., Meshalkina, N.S., Sych, R.A., and Yan, Y.: 2011, *Sol. Phys*. 273, 163.
Aschwanden M., 2004, Physics of the Solar Corona. An Introduction, Springer, Praxis Publishing Ltd, Chichester, UK.
Bárta, M. and Karlický, M.: 2006, *Astron. Astrophys*. 450, 359.
Bouratzis C., Hillaris, A., Alissandrakis, C. E., Preka-Papadema, P., Moussas, X.C., Caroubalos, Tsitsipis, P., and A. Kontogeorgos 2016, *Astron. Astrophys.* 586, A 29.
Breizman, B.N.: 1987, in *Problems in Plasma Theory*, Kadomtsev, B.B. Eds., Energoizdat, Moscow (in Russian), 15, 55.
Chen Bin, T.S.Bastian, D.E.Gary, and Ju Jing, 2011, Spatally and specrally resolved observations of a zebra pattern in solar decimetric radio burst, *Astrophys. J.* 736, 64.
Chernov, G.P.: 1976, *Sov. Astron.* 20, 582.
Chernov, G.P.: 1990, *Sol. Phys*. 130, 75.
Chernov, G.P.: 1997, *Astronomy Letters*, **23**, 827.
Chernov, G.P.: 2005, Recent data on zebra patterns, Astron. Astrophys. 437, 1047-1054.
Chernov, G.P.: 2006, *Space Sci. Rev.* 127, 195.
Chernov, G.P., and Zlobec, P.: 1995, *Solar Phys.* 160, 79.
Chernov, G.P., Fu, Q., Lao, D.B. and Hanaoka, J.: 2001, *Solar Phys.* 201, 153.
Chernov, G.P., 2011, Fine structure of solar radio bursts, Springer ASSL 375, Heidelberg.
Chernov G.P., Sych, R. A., Meshalkina, N. S., Yan, Y. and Tan C. 2012a: *Astron. Astrophys*. 538, A53.
Chernov G.P., 2012b, Сomplex Radio Zebra Patterns escaping from the Solar Corona and New Generation Mechanisms, In: Horizons in World Physics, V. 278, Eds: Albert Reiner, Nova Science Publisher, New-York, Ch. 1, P. 1-75.
   https://www.novapublishers.com/catalog/product_info.php?products_id=30998
Chernov, G.P., Yan, Y.H., and Fu, Q.J.: 2003, *Astron. Astrophys*. 406, 1071.
Chernov, G.P., Yan, Y.H., Fu, Q.J. and Tan Ch.M.: 2005, *Astron. Astrophys*. 437, 1047.
Chernov, G.P., Yan, Y.H, Fu. Q.J.: 2014a, *Research in A&A*, 14, 7, 831.
Chernov, G.P., Fomichev, V.V., Gorgutsa, R. V. Markeev A. K., Sobolev, D. E. Hillaris, A. Alissandrakis, K.: 2014b, *Geomagnetism and Aeronomy,* 54, 406.
Chernov, G.P., Fomichev, V.V., Tan, B.L., Yan, Y.H., Tan, Ch.M. and Fu, Q.J.: 2015, *Solar Phys.* 290, 95.
Chernov, G.P., Fomichev, V.V., Tan B.L., Yan, Y.H, Tan Ch.M., Fu. Q.J.: 2016, *Res. Astron. Astrophys*. 16, 28.
Fomichev, V.V. and Fainshtein, S.M.: 1988, *Sov. Astron.* 32, 552.
Gao, G. Dong, M., L. Wu, N. Lin, J.: 2014, *New Astronomy,* 30, 68.
Gladd, N.T.: 1983, *Phys. Fluids*, 26, 974.
Hankins, T. H., & Eilek, J. A. 2007, ApJ, 670, 693.
Huang G., and Lin J.: 2006, ApJ, 639, L99.
Huang, G.L., and Tan, B.L.:2012, *Astrophys. J.* 745, 186.
Kaneda, K., Misawa, H. Iwai, K. Tsuchiya, F. and Obara, T.: 2015, *Astrophys. J. L.* 808, L45.

Karlický, M., Barta, M., Jiricka, K. et al.: 2001, *Astron. Astrophys.* 375, 638.
Karlický, M., Meszarosova, H., & Jelinek, P.: 2013, *Astron. Astrophys.* 550, A1
Karlický, M.: 2013, *Astron. Astrophys.* 552, A90.
Karlický, M. and Yasnov, L.V.: 2015, *Astron. Astrophys.* 581_A115.
Kuijpers, J.: 1975, *Collective wave-particle interactions in solar type IV radio source*, Ph.D Thesis, Utrecht University.
Kuijpers, J.: 1980, in M.R.Kundu and T.E.Gergely (eds), *Theory of type IV dm Bursts,* Radio Physics of the Sun, p. 341.
Kuznetsov, A. A.: 2005, *Astron. Astrophys.* 438, 341.
Kuznetsov, A.A.: 2006, *Solar Phys.* 237, 153.
Kuznetsov, A. A. 2007, Astron. Letters., 33, 319.
Kuznetsov, A.A., and Tsap, Yu.T.: 2007, *Solar Phys.* 241, 127.
LaBelle J., Treumann R.A., Yoon P.H., Karlický M.: 2003, *Astrophys. J.* 593, 1195.
Laptuhov, A.I. and Chernov, G.P.: 2006, *Plasma Phys. Rep.* 32, 866.
Laptuhov, A.I. and Chernov, G.P.: 2009, *Plasma Phys. Rep.* 35, 160.
Laptuhov, A.I. and Chernov, G.P.: 2012, *Plasma Phys. Rep.* 38, 613.
Ledenev, V. G., Yan, Y., and Fu, Q.: 2006, *Sol. Phys.*, 233, 129.
Melnik, V.N., Rucker, H.O, and Konovalenko, A.A.: 2008, Solar Physics Research Trends (ed. P. Wang) Nova Science Publisher, NY, Ch.8, 287.
Mollwo, L.: 1983, *Solar Phys.* 83, 305.
Mollwo, L.: 1988, *Solar Phys.* 116, 323.
Selhorst, C. L., Silva-Válio, A., and Costa, J. E. R. 2008, A&A, 488, 1079
Slottje, C.: 1981, *Atlas of fine Structures of Dynamic Spectra of Solar Type IV-dm and Some Type II Bursts,* Utrecht Observatory.
Stepanov, A.V., Tsap, Yu.T.: 1999, *Astron. Rep.* **43**, 838.
Tan B.L., Yan Y.H., Tan C.M., Sych R.A., Gao G.N., 2012, Microwave zebra pattern Structures in the X2.2 solar flare on Feb. 15, 2011, *Astrophys. J.* 744, 166-183.
Tan B.L., Tan C.M., 2012, Astrophys. J. 749, 28-35.
Tan, B. L., Tan, C. M., Zhang, Y., Mészárosová, H., Karlický, M., 2014a, *Astrophys. J.*, 780, 129.
Tan, B. L., Tan, C. M., Zhang, Y., Huang, J. Mészárosová, H., Karlický, M., Yan, Y.2014b, *Astrophys. J.*, 790, 151.
Treumann, R. A., Nakamura, R., Baumjohann, W.: 2011, Ann. Geophys., 29, 1673.
Tsang, K.T.: 1984, *Phys. Fluids*, 27, 1659.
Viktorov, M., Mansfeld, D., and Golubev, S.: 2015a, EPL, 109 (2015) 65002.
Viktorov, M. E. Golubev, S. V. Zaitsev, V. V. and Mansfeld, D.A.: 2015b, *Radiophysics and Quantum Electronics, 57,* 849.
Winglee, R.M. and Dulk G.A.: 1986, *Astrophys. J.* 307, 808.
Yasnov, L.V., and Karlický, M.: 2015, *Solar. Phys*, 290, 2001.
Yan, Y. H., Zhang, J., Wang, W., et al.: 2009, Earth Moon and Planets, 104, 97.
Yan, Y., Wang, W., Liu, F., et al.: 2013, in IAU Symposium, 294, eds. A. G. Kosovichev, E. de Gouveia Dal Pino, and Y. Yan, 489.
Yu, S. J., Yan, Y. H.,and Tan, B. L.: 2012, *Astrophys. J.*, 761, 136.
Yu, S. J., Nakariakov, V. M., Selzer, L. A., Tan, B. L., & Yan, Y. H.: 2013, *Astrophys. J.* 777, 159.
Zheleznyakov V.V.: 1995, *Radiation in Astrophysical Plasma*, Kluwer Academic, Dordrecht (in Russ. Izdat. Nauka, Moscow, 1977).
Zheleznykov, V.V., and Zlotnik, E.Ya.: 1975a, *Solar. Phys.* 44, 447.
Zheleznykov, V.V., and Zlotnik, E.Ya.: 1975b, *Solar. Phys.* 44, 461.
Zlotnik, E.Ya.: 1976, *Izv. VUZov Radiofizika,* 19, 481.
Zlotnik E.Ya.: 2009, Origin of zebra pattern in type IV solar radio emission, Cent. Eur. Astrophys. Bull. 1, 281-298.
Zlotnik, E.Ya.: 2010, *Solar-Terrestrial Physics, Collected articles*, Sibirian Depatement of Russian Academy of Sciences Eds, (in Russian), 16, 49.
Zlotnik E.Ya.: 2013, *Solar. Phys.* 284, 579.
Zlotnik E.Ya, Zaitsev V.V., Altyntsev, A.T.: 2014, *Solar. Phys.* 289, 233.

Zlotnik E.Ya, Zaitsev V.V., Aurass H., Hofmann A.: 2003, Solar type IV burst spectral fine structures- Part II- Source model, *Astron. Astrophys*. 410, 1011.

Zlotnik E.Ya., Zaitsev V.V., Melnik V.N, Konovalenko A.A. and Dorovskyy V.V.: 2015, *Solar. Phys.* 290, 2013.